\documentclass{article}
\usepackage{amsfonts,amsmath,axodraw,pstricks,color,graphicx}
\begin{document}

\renewcommand{\baselinestretch}{1.2}

\newcommand{\Mvariable}{{}}
\newcommand{\be}{\begin{equation}}
\newcommand{\ee}{\end{equation}}
\newcommand{\bs}{\begin{split}}
\newcommand{\es}{\end{split}}
\newcommand{\LA}{\langle}
\newcommand{\RA}{\rangle}
\newcommand{\lar}{\leftarrow}
\newcommand{\rar}{\rightarrow}
\newcommand{\bea}{\begin{eqnarray}}
\newcommand{\ol}{\overline}
\newcommand{\eea}{\end{eqnarray}}
\newcommand{\nn}{\nonumber}
\newcommand{\ola}{\overleftarrow}
\newcommand{\ora}{\overrightarrow}
\newcommand{\p}{\partial}
\newcommand{\half}{\frac{1}{2}}
\newcommand{\m}{\mu}
\newcommand{\n}{\nu}
\newcommand{\la}{\lambda}
\newcommand{\gra}{\alpha}
\newcommand{\gre}{\epsilon}
\newcommand{\g}{\gamma}
\newcommand{\de}{\delta}
\newcommand{\ta}{\theta}
\newcommand{\nc}{\Theta}
\newcommand{\vt}{\vartheta}
\newcommand{\vf}{\varphi}
\newcommand{\ca}{c_{\alpha}}
\newcommand{\sa}{s_{\alpha}}
\newcommand{\ct}{c_{\theta}}
\newcommand{\st}{s_{\theta}}
\newcommand{\hgamma}{\hat{\gamma}}
\newcommand{\hbeta}{\hat{\beta}}
\newcommand{\hnc}{\hat{\Theta}}
\newcommand{\hrho}{\hat{\rho}}
\newcommand{\hb}{\hat{b}}
\newcommand{\tnc}{\widetilde{\Theta}}
\newcommand{\tB}{\widetilde{B}}
\newcommand{\tphi}{\widetilde{\phi}}
\newcommand{\diff}{\mathrm{d}}
\newcommand{\e}{\mathrm{e}}
\newcommand{\rmS}{\mathrm{S}}
\newcommand{\AdS}{$\mathrm{AdS}_5 \times \mathrm{S}^5$}
\newcommand{\Ncal}{\mathcal{N}}
\newcommand{\Gcal}{\mathcal{G}}
\newcommand{\Ocal}{\mathcal{O}}
\newcommand{\Wcal}{\mathcal{W}}
\newcommand{\Tr}{\mathrm{Tr}}
\newcommand{\Tcal}{\mathcal{T}}

\newcommand{\Acal}{{\mathcal A}}
\newcommand{\Rcal}{{\mathcal R}}
\newcommand{\Dcal}{{\mathcal D}}
\newcommand{\Mcal}{{\mathcal M}}
\newcommand{\Lcal}{{\mathcal L}}
\newcommand{\Scal}{{\mathcal S}}
\newcommand{\Bcal}{\mathcal{B}}
\newcommand{\Ccal}{\mathcal{C}}
\newcommand{\Vcal}{\mathcal{V}}
\newcommand{\Zb}{\overline{Z}}
\newcommand{\Urm}{{\mathrm U}}
\newcommand{\Srm}{{\mathrm S}}
\newcommand{\SO}{\mathrm{SO}}
\newcommand{\Sp}{\mathrm{Sp}}
\newcommand{\SU}{\mathrm{SU}}

\hyphenation{} \hyphenation{} \hyphenation{}

%\DeclareSymbolFont{AMSa}{U}{msa}{m}{n}
%\DeclareSymbolFont{AMSb}{U}{msb}{m}{n}
%\DeclareMathSymbol{\fieldR}{\mathalpha}{AMSb}{"52}
%\newif\ifpdf
%\ifx\pdfoutput\uundefined
%\pdffalse % we are not running PDFLaTeX
%\else
%\pdfoutput=1 % we are running PDFLaTeX
%\pdftrue \fi

\begin{titlepage}
\begin{center}
\hfill {\tt YITP-SB-06-60}\\
\vskip 20mm

{\Large {\bf Marginal Deformations of $\Ncal=4$ SYM and Open vs. Closed String Parameters}}

\vskip 10mm

{\bf Manuela Kulaxizi}

\vskip 4mm {\em C. N. Yang Institute for Theoretical
Physics}\\
{\em State University of New York at Stony Brook}\\
{\em Stony Brook, NY 11794-3840, USA}\\
[2mm] E-mail: {\tt kulaxizi@insti.physics.sunysb.edu}
\end{center}
\vskip 1in

\begin{center} {\bf ABSTRACT }\end{center}

\vskip 4pt  
We make precise the connection between the generic Leigh--Strassler deformation of $\Ncal=4$ SYM
and noncommutativity. We construct an appropriate noncommutativity matrix, which turns out to define a 
nonassociative deformation. Viewing this noncommutativity matrix as part of the set of open string data which 
characterize the deformation and mapping them to the closed string data (e.g. metric and B--field), we are able to 
construct the gravity dual and the the correponding deformed flat space geometry up to third order in the deformation parameter $\rho$.

\begin{quotation}
\noindent

\end{quotation}
\vfill \flushleft{December 2006}
\end{titlepage}
\eject

%%%%%%%%%%%%%%%%%%%%%%%%%%
\section{Introduction}
%%%%%%%%%%%%%%%%%%%%%%%%%%

The AdS/CFT correspondence \cite{Maldacena9711,Gubseretal98,Witten9802} offers an equivalence between 
gauge theory and gravity. In its original form, relates superconformal $\Ncal=4$ SU(N) Super Yang--Mills 
to closed string theory on \AdS with N units of RR--flux. While closed strings on nontrivial backgrounds
with RR--fluxes are still in many ways intractable, their low energy description in terms of 
supergravity is not. From the gauge theory point of view, this limit corresponds to large N 
and strong t'Hooft coupling $\lambda$. This makes the correspondence extremely useful in that it provides
a window into understanding the physics of gauge theories in a region that is otherwise difficult 
to explore. By now the original proposal has been greatly extended covering gauge theories with less
amount of supersymmetry and/or a running coupling constant \cite{KlebanovWitten, KlebanovStrassler, MaldacenaNunez, KachruSilverstein, Lawrenceetal, Gauntlettetal0402}.

%Despite however the progress made thus far the natural question of how to specify the gravitational %background
%corresponding to a given gauge theory, is still far from being understood 
%CITATIONS
%In this article we will attempt to give an alternative point of view on this matter while investigating
%marginal deformations of the $\Ncal=4$ theory. 

The simplest extensions of the original AdS/CFT proposal arise by considering
supersymmetry preserving deformations of the $\Ncal=4$ SYM theory; exactly marginal and/or relevant deformations. Naively, the gravity dual backgrounds of the former would be more accessible 
than those of the latter. It turns out however that the opposite is true.
In fact the gravity duals of a class of supersymmetric mass deformations
were discovered quite early on; see for example \cite{PilchWarner0002, PilchWarner0004, Halmagyietal} and references therein. 
The main reason is that these backgrounds can be analyzed using the truncation to five--dimensional supergravity, something which is not possible for marginal deformations of the $\Ncal=4$ theory. Actualy it was only fairly recently that the authors of \cite{LuninMaldacena} succeeded in constructing the corresponding backgrounds for a subclass of these latter theories\footnote{For a general approach on how to find the gravity dual a description of a given gauge theory see for instance \cite{SkenderisTaylor0603,SkenderisTaylor0604, Gauntlettetal0402,Gauntlett:2004hs,Gauntlettetal0510,Gauntlett:2006vf, Wijnholt:2005mp} and references therein.}. 

Marginal deformations of $\Ncal=4$ SYM preserve $\Ncal=1$ supersymmetry and are mainly described by 
two parameters, denoted as $\beta$ and $\rho$, in addition to the gauge coupling $g_{YM}$. 
In \cite{LuninMaldacena} Lunin and Maldacena discovered the geometry dual to the 
$\beta$--deformed theory i.e. when $\rho=0$. In this case apart from the $U(1)$ R symmetry, the theory preserves additional two $U(1)$ global symmetries. These symmetries played a significant part in the construction of the new solution.  
When $\rho\neq 0$ however, the theory does not preserve any continuous symmetries other than 
the U(1) R--symmetry (only a discrete $Z_3\times Z_3$ symmetry) and the problem of finding the dual gravitational background has resisted solution.

In this note, we revisit the question of how to obtain the gravitational dual of the $\rho$--deformation. 
The starting point is to make precise the description of the deformation in the dual gravity theory as a non-commutative deformation of the transverse space. The relation between exactly marginal deformations of $\Ncal=4$ SYM and non-commutativity was actually established early on \cite{Berensteinetal00} 
(see also \cite{KulaxiziZoubos04,KulaxiziZoubos05} for a non-commutative description at weak coupling). For the $\beta$--deformation it was only made explicit recently, in \cite{LuninMaldacena}. 
Here we attempt to make the relation to non-commutativity explicit for the $\rho$ --deformation as well. In particular, we construct the non-commutativity matrix $\nc$ which practically realizes the deformation. The $(2,0)$ and $(0,2)$ parts of the non-comuutativity matrix are easily obtained from the F-term constraints. To specify the $(1,1)$ components, we follow the discussion in \cite{Kulaxizi0610}. We first consider all possibilities allowed by the global discrete $Z_2\times Z_3$ symmetries. However, symmetries do not adequately constrain the form of $\nc$. To determine it completely we tansform it to spherical coordinates and require that it be real, transverse and independent of the radial direction. These restrictions are imposed upon us from the exact marginality of the deformation and specify $\nc$ completely.

It turns out that we can perform a rather non-trivial check on the proposed non-commutativity matrix. 
%We then confirm the validity of the non-commutativity parameter thus constructed by a rather non-trivial %check. 
There exist points in the Leigh-Strassler deformation space parametrized by $(\beta,\rho)$, which are related to each other by a field redefinition. For instance, the $\Ncal=4$ SYM theory deformed by $(\beta_1=0,\rho_1\in R)$ is equivalent to the same theory deformed by $(\beta_2,\rho_2)$ such that $\rho_2={i\rho_1\over \sqrt{3}}={i\beta_2\over 2}$. 
Clearly, the associated non-commutativity matrices $\nc_1,\,\nc_2$ should also be related if they correctly describe the deformations. We find that this is indeed the case; a simple coordinate transformation takes us from $\nc_1$ to $\nc_2$ confirming the equivalence of the two theories in this description. 
A disconcerting fact about $\nc$ is that it does not satisfy the associativity condition. Hence, we do not have a star--product formulation which would enable us to express the superpotential of the deformed theory in terms of the parent $\Ncal=4$ theory; (this was succesfully done for the case of the $\beta$--deformation \cite{LuninMaldacena} -- see also \cite{Khoze}).

%It is interesting to check whether the associated non-commutativity matrices are also related
%We find that the non-commutativity matrices associated to these two deformations are related by a %coordinate transformation, showing that the two theories are equivalent in this description as well.
%for which 
%CHECK of matrix + footnote on Costas work + mention non-associativity.

One may wonder whether we can use the knowledge of the non-commutativity matrix to obtain information on the gravity dual of the theory. A way to address this question, is perhaps to consider the non-commutativity matrix as part of the gauge theory or open string theory data (metric+non-commutativity) which characterize the deformed theory. One might then hope to obtainsome of the closed string data (metric+B-field) using the relations of Seiberg and Witten \cite{SeibergWitten9908} (see also \cite{Schomerus} ).

We examine this possibility by directly employing the Seiberg-Witten relations for the $\rho$--deformation.
Unfortunately, the non-commutativity matrix which describes the $\rho$--deformation does not satisfy the associativity condition. Most likely the Seiberg--Witten relations are not
valid for non-associative deformation parameters. Nevertheless, nonassociativity is
a second--order in $\rho$ effect, so one may hope to at least obtain a correct result up to third order in the deformation parameter. 
Indeed we find that the set of closed string data deduced from the Sieberg-Witten relations, {\it i.e.} metric, dilaton and B--field, satify the field equations of supergravity up to third order in the deformation parameter $\rho$. Since all the necessary symmetries are built in, we expect that this deformed flat space geometry presents, up to third order in the deformation parameter, the  background where once D--branes are immersed and the near horizon limit is taken, the AdS dual geometry will be recovered.

Given this result, we consider the effective action of the $\rho$--deformed gauge theory, obtained by giving a vacuum expectation value to one of the scalars and integrating out the massive fields. According to \cite{Maldacena9705,Maldacena9709,Tseytlin:1999dj, MetsaevTseytlin,ChepelevTseytlin,Balasubramanianetal}, the leading IR large N part of this action should coincide with the DBI action for a D3--brane immersed in the dual background. We observe that in the case of the $\beta$--deformed gauge theory, the corresponding DBI action
is characterized by the open string data $(\Gcal_{AdS_{5}\times \Srm^{5}},\nc)$ and that
the associated NS--NS closed string fields $(g,B)$ are part of the \emph{exact}
Lunin--Maldacena solution. This is not surprising. Indeed, the Lagrangian description 
of this theory can be given in terms of the $\Ncal=4$ Lagrangian with the product of matter fields 
replaced by a star product of the Moyal type. Subsequently, all amplitudes in the planar limit can 
be shown \cite{Khoze} to be proportional up to a phase to their $\Ncal=4$ counterparts. Then the
open string data $(\Gcal_{AdS_{5}\times \Srm^{5}},\nc=0)$ of the $\Ncal=4$ SYM theory are naturally promoted to the set $(\Gcal_{AdS_{5}\times \Srm^{5}},\nc)$. Can something similar occur for the 
$\rho$--deformation? 

Non-associativity again creates a potential problem: planar equivalence with the parent $\Ncal=4$ theory using a star--product is far from obvious. Nonetheless, nonassociativity is
a second--order in $\rho$ effect, so we can safely assume that $(\Gcal_{AdS_{5}\times \Srm^{5}},\nc)$ describe the deformation up to this order in the deformation parameter.  
We then map the open string fields to the closed ones using the Seiberg-Witten relations and obtain the gravity dual of the $\rho$--deformed gauge theory up to third order in $\rho$.

The structure of this paper is the following: In section 2 we review some known facts about marginal
deformations of the $\Ncal=4$ theory and their gravity duals. In addition, we explore some special 
points in the deformation space for which the general theory with $\beta\neq 0$ and $\rho\neq 0$
is equivalent to an exactly marginal deformation with either $\widetilde{\beta}=0$ or $\widetilde{\rho}=0$.
In section 3, we use the logic outline in \cite{Kulaxizi0610} and determine the noncommutativity matrix for the $\rho$--deformation. Viewing $\nc$ as part of the open string data which describe the deformation, we use the Seiberg-Witten relations to find the corresponding closed string data $(g,B)$. This procedure
is illustrated in section 4 where we derive the $\rho$--deformed flat space geometry up to third order 
in the deformation parameter. In section 5 we proceed with considerations on the DBI action which 
provide us with the gravity dual of the $\rho$--deformed theory to the same order. 
We conclude in section 6.
%Section 6, contains an extensive discussion on the ideas set forth in this note, whereas conclusions and 
%possible extensions are given in section 7.

%%%%%%%%%%%%%%%%%%%%%%%
\section{The Leigh Strassler Deformation}
%%%%%%%%%%%%%%%%%%%%%%%

Not long after it was realized that $\Ncal=4$ SU(N) Super Yang Mills theory is finite (see e.g.
\cite{Sohnius85} for an account), it became clear that it might not be the only four dimensional 
theory with that property \cite{Jones86, ParkesWest84, JonesMezincescu84, ParkesWest85, Grisaruetal85}. 
It was however almost ten years later, when Leigh and Strassler undertook a systematic study
of marginal deformations of $\Ncal=4$ and indeed showed that there exists a whole class of
$\Ncal=1$ supersymmetric gauge theories satisfying both the requirements for conformal invariance
and finiteness \cite{LeighStrassler95}. More precisely, they showed that the $\Ncal=4$ theory
admits a three--complex--parameter family of marginal deformations which preserve $\Ncal=1$
supersymmetry and are described by the following superpotential:
\begin{equation} \label{Superpotential}
\Wcal=i h \mathrm{Tr} \left[ \left(e^{i\beta}\Phi_1\Phi_2\Phi_3-e^{-i\beta}\Phi_1\Phi_3\Phi_2\right)
+\rho \left(\Phi_1^3+\Phi_2^3+\Phi_3^3\right)\right]
\end{equation}
where $\Phi^{I}$ with $I=1,2,3$ are the three chiral superfields of the theory.
Together with the gauge coupling $g_{YM}$, the complex parameters ($h, \beta, \rho$) 
that appear in the superpotential constitute the four couplings of the theory.

While it is clear at the classical level that these deformations are marginal --- since
all operators of the component Lagrangian have classical mass dimension equal to four ---
this is not necessarily true quantum mechanically. Leigh and Strassler realized that by using the 
constraints of $\Ncal=1$ supersymmetry and the exact NSVZ beta--functions 
\cite{Novikovetal, ShifmanVainshtein86, ShifmanVainshtein91} 
written in terms of the various amonalous dimensions of the theory, it was possible to express 
the conditions for conformal invariance of the quantum theory, through linearly dependent equations 
which were therefore likely to have nontrivial solutions.
In this way, they were able to demonstrate that the deformation of (\ref{Superpotential}) is truly
marginal at the quantum level, so long as the four couplings of the theory satisfy a single
complex constraint $\gamma(g_{YM},h,\beta,\rho)=0$. In other words, there exists a  
three--complex--dimensional surface $\gamma(g_{YM},h,\beta,\rho)=0$ in the space of
couplings, where both beta functions and anomalous dimensions vanish and thus the $\Ncal=1$
gauge theories mentioned above are indeed conformally invariant. In general, the function 
$\gamma$ is not known beyond two--loops \cite{AharonyRazamat,Razamat:2002tm,NiarchosPrezas02, FreedmanGursoy, Maurietal05} 
in perturbation theory, where it reads:  
\be\label{Fullcondition}
|h|^2 \left[\frac{1}{2} \left(|q|^2+\frac{1}{|q|^2}\right)-\frac{1}{N^2}\left|q-\frac{1}{q}\right|^2 
+|\rho |^2\left(\frac{N^2-4}{2 N^2}\right) \right]= g_{YM}^2
\ee
with q defined as $q=\mathrm{e}^{i\beta}$ and N the number of colours of the gauge theory.

For the $\beta$--deformed gauge theory, \emph{i.e.}, 
obtained by setting $\rho=0$ in the superpotential of equation (\ref{Superpotential}),
the Leigh--Strassler constraint at two loops can be written as:
\be\label{BetaCondition}
|h|^2 \left[\frac{1}{2} \left(|q|^2+\frac{1}{|q|^2}\right)-\frac{1}{N^2}\left|q-\frac{1}{q}\right|^2 
\right]= g_{YM}^2
\ee
In this case, one immediately notices that when $\beta=\beta_{\mathbb{R}}\in\mathbb{R}$ therefore $|q|=1$,
 (\ref{BetaCondition}) reduces to:
\be
|h|^2 \left[1-\frac{1}{N^2}\left|q-\frac{1}{q}\right|^2 
\right]=|h|^2 \left(1- \frac{4}{N^2} \sin^2{\beta_{\mathbb{R}}}\right) = g_{YM}^2
\ee
which in the large N limit yields: $|h|^2=g_{YM}^2$.
Despite the fact that this result was obtained from the two--loop expression of
the conformal invariance condition, it has been shown to be true to all orders 
in perturbation theory at the planar limit \cite{Khoze} (see also \cite{KuzenkoTseytlin,Maurietal05}). 
Actually the author of \cite{Khoze} went even further and showed that all planar 
amplitudes in the $\beta=\beta_{\mathbb{R}}\in\mathbb{R}$ theory are proportional to their $\Ncal=4$ 
counterparts, thus explicitly proving finiteness and conformal invariance. The proof 
made use of an existing proposal \cite{LuninMaldacena} for an equivalent ''noncommutative'' 
realization of the theory. 
For the more general case of complex $\beta=\beta_{\mathbb{R}}+i\beta_{\mathbb{I}}$, 
equation (\ref{BetaCondition}) in the planar limit reads:
\be
\frac{1}{2} |h|^2 \left(|q|^2+\frac{1}{|q|^2}\right)
=|h|^2 \cosh{(2\beta_{\mathbb{I}})} =g_{YM}^2
\ee
It is then evident that the coupling constant h receives corrections with respect to its
$\Ncal=4$ SYM value. Nevertheless, diagrammatic analysis \cite{Khoze} showed that all planar amplitudes 
with external gluons are equal to those of the $\Ncal=4$ theory up to a five--loop level. 
To this order and beyond, it is most likely that the planar equivalence between the parent 
theory and its deformation will break down.
For (more) recent investigation on $\beta$--deformations from the gauge theory point of view 
see \cite{FreedmanGursoy,Maurietal05,Rossi:2005mr,Rossi:2006mu,Penatietal,Maurietal06,Elmetti:2006gr}.

Special points along the deformation occur when $q=\mathrm{e}^{i\beta}$ is a root of unity.
These points have been studied early on \cite{BerensteinLeigh00,Berensteinetal00} 
with a dual interpretation as orbifolds with discrete torsion. 
The marginally deformed theories have been further explored in
\cite{Doreyetal0210,Dorey03,Dorey04}, and several remarkable
properties have been demonstrated. In particular it was shown that
as expected, the S--duality of $\Ncal=4$ extends to their space of vacua, 
and that, again for special values of $\beta$, there are also new Higgs branches 
on moduli space. These are mapped by S--duality to completely new, confining branches 
which appear only at the quantum level. Furthermore, at large $N$ the
Higgs and confining branches can be argued to be described by Little
String Theory \cite{Dorey04}. Finally, the possibility of an underlying integrable
structure for the deformed theories in analogy with $\Ncal=4$ SYM,  was investigated 
at special values of the deformation parameter in \cite{BerensteinCherkis04,Bundzik:2005zg} 
and for generic $\beta$ in \cite{BeisertRoiban,Frolov,McLoughlinSwanson,Aldayetal05}.

%%%%%%%%%%%%%%%%%%%%%%%%%%%%%%%%%%%%%%%%%%%%%%%%%%%%%%%%%%%%%%%%%%%%%%%%%%%%%%%%%%%%%%%%%%%%%%%

%%%%%%%%%%%%%%%%%%%%%%%%%%%%%%%%%%%%%%%%%%%%%%%%%%%%%%%%%%%%%%%%%%%%%%%%%%%%%%%%%%%%%%%%%%%%%%%
\subsection{Marginal deformations and gauge/gravity duality}

A natural place to explore theories that arise as marginal deformations 
of $\Ncal=4$ SU(N) SYM is the AdS/CFT correspondence where the strong coupling regime of
the undeformed theory is realized as weakly coupled supergravity on \AdS. Due to
superconformal symmetry, the dual gravitational description of these theories is expected 
to be of the form: $\mathrm{AdS}_{5}\times \tilde{\mathrm{S}}^5$ with $\tilde{\mathrm{S}}^5$
a sphere deformed by the presence of additional NS--NS and RR fluxes.
Indeed in \cite{Aharonyetal02}, where the dual background was constructed to second order
in the deformation parameters, it was shown that apart from the already present five--form flux
one should also turn on (complexified) three--form flux $G_{(3)}$ along the $\Srm^5$. 

Essential progress however in this direction was only recently achieved through the work
of Lunin and Maldacena \cite{LuninMaldacena}. The authors of \cite{LuninMaldacena} 
succeeded in finding the \emph{exact} gravity dual of the $\beta$--deformed gauge theory.

In this case, apart from the $U(1)_{R}$ R--symmetry the theory preserves two global $U(1)$s, 
which act on the superfields in the following way:
\be\label{U1}
\begin{split}
U(1)_{1}&:\quad \left(\Phi_{1},\Phi_{2},\Phi_{3}\right)\rightarrow(\Phi_{1},
\e^{i \alpha_{1}}\Phi_{2},\e^{-i\alpha_{1}}\Phi_{3})\\
U(1)_{2}&:\quad \left(\Phi_{1},\Phi_{2},\Phi_{3}\right)\rightarrow(\e^{-i \alpha_{2}}\Phi_{1},
\e^{i \alpha_{2}}\Phi_{2},\Phi_{3})
\end{split}
\ee
%Furthermore, they proposed a method that led to the explicit construction
%of several new supergravity backgrounds. 
The main idea underlying the solution generating technique proposed in \cite{LuninMaldacena}, 
was the natural expectation that the two U(1) symmetries preserved 
by the deformation would be realized geometrically in the dual gravity solution. 
For $\beta=\beta_{\mathbb{R}}\in\mathbb{R}$ their prescription amounts to performing an 
$SL(2,\mathbb{R})$ transformation on the complexified K\"ahler modulus 
$\tau$ of the two torus associated with the U(1) symmetries in question. The specific element of  
$SL(2,\mathbb{R})$ under consideration is: 
$\left(\begin{smallmatrix}
a&b\\
c&d\\
\end{smallmatrix}\right)\equiv
\left(\begin{smallmatrix}
1&0\\ 
c&1\\
\end{smallmatrix}\right)$. 
It is chosen so as to ensure that the new solution will present no singularities 
as long as the original one is non--singular and its sole free parameter $c$ 
is naturally identified with the real deformation parameter $\beta_{\mathbb{R}}$ of 
the gauge theory.

Later on, the method of Lunin and Maldacena was reformulated \cite{Aybike} in terms of the action of a 
T--duality group element on the background matrix $E=g+B$ providing a significantly easier way of 
obtaining the new solutions. In particular, it was shown \cite{Aybike} that one can embed the $SL(2,\mathbb{R})$ 
that acts on the K\"ahler modulus into the T--duality group $O(3,3,\mathbb{R})$
in the following way: 
\be\label{O33LM} 
\Tcal=\begin{pmatrix}
\mathbf{1}& \mathbf{0}\\
\mathbf{\Gamma}& \mathbf{1}\\
\end{pmatrix}\quad \text{where now}\quad 
\mathbf{\Gamma} \equiv \begin{pmatrix}
0 & -\beta_{\mathbb{R}} & \beta_{\mathbb{R}}\\
\beta_{\mathbb{R}} & 0 & -\beta_{\mathbb{R}}\\
-\beta_{\mathbb{R}} & \beta_{\mathbb{R}} & 0\\
\end{pmatrix}
\ee
where $\mathbf{1}$  and  $\mathbf{0}$ represent the $3 \times 3$ identity and zero 
matrices respectively. Suppose then that $E_{0}=g_{0}+B_{0}$ denotes the part of the original 
supergravity background along the U(1) isometry directions which are to be deformed.
Acting on $E_{0}$ with the T--duality group element $\Tcal$ of (\ref{O33LM}) one obtains the 
NS--NS fields of the deformed solution in terms of $E_{0}$ and $\Gamma$ according to:
\be\label{Tduality}
\begin{split}
E&=\frac{1}{E_{0}^{-1}+\mathbf{\Gamma}}\\
e^{2\Phi}&=e^{2\Phi_{0}} \det({1+E_{0}\mathbf{\Gamma}})\equiv e^{2\Phi_{0}} G
\end{split}\ee
The RR-fields of the background can be computed using the T--duality
transformation rules of \cite{Sundell,Bergshoeffetal,Cveticetal,Fukumaetal,Hassan}, 
however the details of this transformation need not concern us here. 
As an example, let us consider ten--dimensional flat space parametrized as:
\be
\diff s^2=-\diff t^2+\sum_{\mu=1}^3 \diff x^{\mu}\diff x_{\mu}+\sum_{i=1}^{3}(\diff r^2_{i}+
r^2_{i}\diff\vf^2_{i})
\ee
In this case $E_{0}$ will contain the components of the flat metric along the polar 
angles $\vf_{i}$. Applying equations (\ref{Tduality}) yields:
\be\label{LMflat}
\begin{split}
\diff s^2&=-\diff t^2+\sum_{\mu=1}^3 \diff x^{\mu}\diff x_{\mu}+\sum_{i=1}^{3}(\diff r^2_{i}+G
r^2_{i}\diff\vf^2_{i})+\beta_{\mathbb{R}} G
r^2_{1}r^2_{2}r^2_{3}\left(\sum_{i=1}^{3}\diff\vf_{i}\right)^2\\
e^{2 \Phi}&=G, \quad G^{-1}=1+\beta_{\mathbb{R}}^2\left(\sum_{i\neq j}r^2_{i}r^2_{j}\right), 
\quad B=\beta_{\mathbb{R}}  G \left(\sum_{i\neq j}r^2_{i}r^2_{j}\diff\vf_{i}\diff\vf_{j}\right)\\
\end{split}
\ee
This is the deformed flat space geometry where by placing D3--branes at the origin
and taking the near horizon limit, one obtains the gravity dual to the $\beta$--deformed
gauge theory. Alternatively, the latter background can be constructed by applying (\ref{Tduality})
on \AdS representing the dual gravitational description of the undeformed parent $\Ncal=4$ theory:
\be\label{LMAdS} 
\begin{split}
\diff s^2=&\mathrm{R}^2(\diff s^2_{\mathrm{AdS}_{5}}+\diff s^2_{5}),\quad \mathrm{where:}\quad
\diff s^2_{5}=\sum_{i}(\diff\m^2_{i}+G
\m^2_{i}\diff\vf^2_{i})+\hbeta G
\m^2_{1}\m^2_{2}\m^2_{3}(\sum_{i}\diff\vf_{i})^2\\
e^{2 \Phi}&=e^{2 \Phi_{0}}G,\quad G^{-1}=1+\hbeta^2(\sum_{i\neq j}\m^2_{i}\m^2_{j}), \quad\quad \hbeta=\mathrm{R}^2\beta_{\mathbb{R}}, \quad\quad
\mathrm{R}^4=4 \pi e^{\Phi_{0}}\mathrm{N} \\
B&=\hbeta R^2 G \left(\sum_{i\neq j}\m^2_{i}\m^2_{j}\diff\vf_{i}\diff\vf_{j}\right) \quad
C_{2}=-\beta_{\mathbb{R}}(16 \pi N)\omega_{1}(\sum_{i}\diff\vf_{i})\\
F_{5}&=(16 \pi N)(\omega_{\mathrm{AdS}_{5}}+G \omega_{\rmS^{5}}),\quad
\omega_{\rmS^{5}}=\diff\omega_{1}\diff\vf_{1}\diff\vf_{2}\diff\vf_{3}, \quad 
\omega_{\mathrm{AdS}_{5}}=\diff\omega_{4}
\end{split}
\ee 
Reformulating the Lunin--Maldacena generating solution technique in terms of the T--dualty group 
action, made especially transparent its relation to similar methods employed in the context
of noncommutative gauge theories\footnote{Evidence relating marginal
deformations and noncommutativity was given earlier both at strong \cite{Berensteinetal00} 
and weak \cite{KulaxiziZoubos04,KulaxiziZoubos05} coupling.}. 

It is easy to see that $\Gamma$ of (\ref{O33LM}) is precisely the
noncommutativity matrix $\nc$ associated with the deformation of the transverse space. In \cite{Kulaxizi0610} the possibility of determining $\nc$ directly from the gauge theory Lagrangian
(and some basic notions of AdS/CFT) was discussed. In particular, it was shown that by promoting the matter fields to coordinates 
$(z^{I},\ol{z}^{\bar{I}})$ 
%parametrizing the space transverse to the D3--brane where the gauge theory lives 
and requiring that $\nc$ should be real, preserve the global symmetries of the theory and respect exact marginality\footnote{More details on this last requirement will be given in the section 3.}, $\nc$ was uniquely fixed to be:
\be\label{ncbetamatrix} 
\nc_{\beta}=a 
\begin{pmatrix} 
0&z_{1}z_{2}&-z_{1}z_{3}&0& -z_{1}\ol{z}_{2}&
z_{1}\ol{z}_{3}\\
-z_{1}z_{2}&0& z_{2}z_{3}&
\ol{z}_{1}z_{2}&0&- z_{2}\ol{z}_{3}\\
 z_{3}z_{1}&- z_{2}z_{3}&0&-
\ol{z}_{1}z_{3}&\ol{z}_{2}z_{3}&0\\
0&-\ol{z}_{1}z_{2}&\ol{z}_{1}z_{3}&0&
\ol{z}_{1}\ol{z}_{2}&-\ol{z}_{1}\ol{z}_{3}\\
z_{1}\ol{z}_{2}&0&-\ol{z}_{2}z_{3}&-
\ol{z}_{1}\ol{z}_{2}&0&\ol{z}_{2}\ol{z}_{3}\\
-\ol{z}_{3}z_{1}&\ol{z}_{3}z_{2}&0& \ol{z}_{1}\ol{z}_{3}&-
\ol{z}_{2}\ol{z}_{3}&0
\end{pmatrix}
\ee
with $a=2\sin{\beta_{\mathbb{R}}}$.
$\nc_{\beta}$ appears to be different from $\Gamma$ of (\ref{O33LM}),
but transforming it to polar coordinates $(r_{i},\vf_{i})$ on $\mathbb{R}^{6}$ one finds that
\be\label{LMThetaPolar} 
\nc_{\beta}=
\begin{pmatrix}
0&0&0&0&0&0\\
0&0&0&0&0&0\\
0&0&0&0&0&0\\
0&0&0&0&-a&a\\
0&0&0&a&0&-a\\
0&0&0&-a&a&0
\end{pmatrix}
\ee
thereby showing that $\nc_{\beta}$ and $\Gamma$ are effectively the same (recall that the Lunin--Maldacena solution is valid for small $\beta_{\mathbb{R}}$ in which case $a\simeq 2\beta_{\mathbb{R}}$).

To obtain the dual background for the general case of complex $\beta$ one needs to perform an additional $SL(2,\mathbb{R})_{s}$ transformation on the solution corresponding to $\beta_{\mathbb{R}}$. 
By $SL(2,\mathbb{R})_{s}$ we denote here the $SL(2,\mathbb{R})$ symmetry of ten dimensional type IIB supergravity
which acts nontrivially on the compexified scalar and two--form fields of the theory.
Being a symmetry of the equations of motion it can be used to generate distinct solutions. 
Subsequent work on the subject of the $\beta$--deformed gauge theories has provided
further checks of the AdS/CFT correspondence 
\cite{Frolovetal0503,Frolovetal0507,ChenKumar,Durnfordetal,GeorgiouKhoze,Hernandezetal} whereas 
generalizations as well as applications of the solution generating technique introduced 
in \cite{LuninMaldacena} were considered in \cite{GursoyNunez,Frolov,Hernandezetal,
AhnVazquezPoritz05,AhnVazquezPoritz06,Rashkovetal}.

%%%%%%%%%%%%%%%%%%%%%%%%%%%%%%%%%%%%%%%%%%%%%%%%%%%%%%%%%%%%%%%%%%%%%%%%%%%%%%%%%%%%%%%%%%%%%%%%%

\subsection{Special points along the general Leigh--Strassler deformation}

In this article we will be mainly interested in the $\rho$--deformed gauge theories. 
In this case --- when $\rho \neq 0$ ---  the theory does not preserve additional U(1) symmetries, 
it is however invariant under a global discrete symmetry  $\mathbb{Z}_{3}\times\mathbb{Z}_{3}$ acting 
on the superfields as:
\be\label{Z3s}
\begin{split}
\mathbb{Z}_{3(1)}&:\quad \left(\Phi_{1},\Phi_{2},\Phi_{3}\right)\rightarrow(\Phi_{3},\Phi_{1},\Phi_{2})\\
\mathbb{Z}_{3(2)}&:\quad \left(\Phi_{1},\Phi_{2},\Phi_{3}\right)\rightarrow(\Phi_{1},\e^{\frac{i2\pi}{3}}\Phi_{2},
\e^{\frac{-i2\pi}{3}}\Phi_{3})
\end{split}
\ee
As previously mentioned, the presence of global U(1)s is crucial in the solution generating 
technique of Lunin and Maldacena which is therefore not applicable here. In fact, the exact gravity dual 
for this case is still unknown. Despite however that the absence of extra continuous symmetries makes the cases 
of $\rho=0$ and $\rho\neq 0$ radically different, there exist special points along the space of couplings 
where the two theories are not only similar but actually equivalent.
%The absense of global continuous symmetries is actually the main distinguishing feature between the cases of
%$\rho=0$ and $\rho\neq 0$.  previously mentioned, the presence of global U(1) symmetries is crucial in the solution generating 
%technique of Lunin and Maldacena, which is therefore not applicable here. In fact, the exact gravity dual 
%for this case is still unknown. In this letter we will use the relation to noncommutativity in order to make 
%progress in this direction, before however we move on to making this relation precise, 
%we would like to consider some special points along the general Leigh--Strassler deformation. 

As first pointed out in \cite{Berensteinetal00} --- see also \cite{Bundzik:2005zg,Bundzik:2006jz}--- 
it is possible to start with either the set $(\beta,\rho)=(\beta,0)$ or $(\beta,\rho)=(0,\rho)$, and via a field 
redefinition reach a point in the deformation space with $(\widetilde{\beta}\neq 0,\widetilde{\rho}\neq 0)$.
The final point will obviously not represent the most general deformation, since the new couplings $\widetilde{\beta}$
and $\widetilde{\rho}$ will be given in terms of the original parameter. In other words, there
will exist a function $f(\widetilde{\beta},\widetilde{\rho})=0$ relating the two.
Furthermore, requiring that the field redefinition be the result of a unitary transformation imposes 
a restriction on the original value of the coupling; be it $\beta$ or $\rho$.
In particular, suppose that we consider the marginally deformed theory at the point
$(\beta,\rho=0)$ and then take: 
\be\label{Fieldredefinition1}
\begin{pmatrix}
\Phi_{1}\\
\Phi_{2}\\
\Phi_{3}\\
\end{pmatrix}\rar
\begin{pmatrix}
A&A&A\\
B&\omega B&\omega^2 B\\
C&\omega^2 C&\omega C\\
\end{pmatrix} 
\begin{pmatrix}
\Phi_{1}\\
\Phi_{2}\\
\Phi_{3}\\
\end{pmatrix}
\ee
with $\Phi_{I}$ the three chiral superfields and $\omega=\mathrm{e}^{i 2\pi/3}$ the third root of unity. 
Note here that since the deformation enters only in the superpotential, it suffices to consider transformations 
that affect the chiral fields independently from the antichiral ones. In other words, we do not expect 
mixing between holomorphic and antiholomorphic pieces. If we furthermore impose the following 
conditions on the free parameters A,B,C: $|A|=|B|=|C|=\frac{1}{\sqrt{3}}$ and 
$ABC=\pm\frac{i \lambda}{3\sqrt{1+2\cos{2\beta}}}$ with $\lambda\in\mathbb{C}$, 
we find that the original $\beta$--deformed gauge theory 
is equivalent to the marginally deformed $\Ncal=4$ SYM theory with coupling constants:
\be\label{betarhonew}
\widetilde{\rho}=\pm\frac{2 \sin{\beta}}{3\sqrt{1+2\cos{2\beta}}} \quad \text{and}\quad 
\mathrm{e}^{i\widetilde{\beta}}=\pm \frac{2 \cos{(\beta-\frac{\pi}{6})}}{\sqrt{1+2\cos{2\beta}}}
\ee
provided that $\beta=\beta_{\mathbb{R}}+i\beta_{\mathbb{I}}$ satisfies the following equation:
\be\label{betapoint}
4 \cos{2 \beta_{\mathbb{R}}} \cos{2 \beta_{\mathbb{I}}}+
4\cos^2{2 \beta_{\mathbb{R}}}+4\cos^2{2 \beta_{\mathbb{I}}}-3\left(1+3 \lambda\right)=0
\ee
Solutions to (\ref{betapoint}) define special regions in the coupling constant space where
the Leigh--Strassler theory with generic $\beta$ and $\rho=0$ is equivalent to a theory
with both $\widetilde{\beta}$ and $\widetilde{\rho}$ nonvanishing but constrained to
satisfy a specific relation dictated from (\ref{betarhonew}).
It is worth remarking here that there is no solution of (\ref{betarhonew}) and (\ref{betapoint})
for which \emph{both} $\beta$ and $\widetilde{\beta}$ are real. This is particularly interesting,
because it is only for the $\beta$--deformed gauge theory with $\beta=\beta_{\mathbb{R}}\in\mathbb{R}$
that a precise connection with noncommutativity is possible.
It is natural to wonder whether distinct unitary field redefinitions of a type similar 
to (\ref{Fieldredefinition1}) could take us from different $\beta$'s to different
$\widetilde{\beta}$ and $\widetilde{\rho}$. It is however not hard to deduce that up to a phase in 
$\widetilde{\rho}$ --- which can be reabsorbed in the definition of the coupling constant h --- 
and a sign in $\widetilde{\beta}$, all such unitary transformations share the same starting point 
(\ref{betapoint}) and lead to the same theory (\ref{betarhonew}).   
  
In an analogous manner one can find specific values of $\rho$ for which the theory
with $\beta=0$ is equivalent to another one with both couplings $\widetilde{\beta}$ and 
$\widetilde{\rho}$ turned on. Detailed analysis in this case shows in fact that such a mapping is 
possible for \emph{any} original value of $\rho$ with parameters $\widetilde{\rho}$ and $\widetilde{\beta}$ 
given by:
\be
\widetilde{\rho}^2=-\frac{\rho^2}{\rho^2+3},\quad\text{and}\quad 
\sin^2{\widetilde{\beta}}=-\widetilde{\rho}^2=\frac{\rho^2}{\rho^2+3}
\ee
The precise field redefinition through which this is achieved, is of the form of (\ref{Fieldredefinition1}):
\be\label{Fieldredefinition2}
\begin{pmatrix}
\Phi_{1}\\
\Phi_{2}\\
\Phi_{3}\\
\end{pmatrix}\rar
\frac{1}{\sqrt{3}}
\begin{pmatrix}
1&1&1\\
1&\omega &\omega^2 \\
1&\omega^2 &\omega \\
\end{pmatrix} 
\begin{pmatrix}
\Phi_{1}\\
\Phi_{2}\\
\Phi_{3}\\
\end{pmatrix}
\ee
Note here again that $\widetilde{\beta}=\widetilde{\beta}_{\mathbb{R}}\in\mathbb{R}$ if and only if $\rho\in\mathbb{R}$
which implies that $\widetilde{\rho}\in\mathbb{I}$. If one additionaly assumes that $\widetilde{\beta}_{\mathbb{R}}\in\mathbb{R}\ll 1$ then the deformed theory with $\beta=0$ and $\rho=q_{1}\in\mathbb{R}$
is equivalent to a theory with $2\sin{\widetilde{\beta}}=\pm 2\frac{q_{1}}{\sqrt{3}}$ and $\widetilde{\rho}=\pm i\frac{q_{1}}{\sqrt{3}}\in\mathbb{I}$. In section 3, we will see that this particular point in the deformation space
naturally shows up in the noncommutative description of the moduli space. This will provide 
us with a non--trivial check on the consistency of the noncommutative interpretation.  

So far we have looked at special points in the space of couplings 
which can be studied at the level of the gauge theory lagrangian. 
There are however a couple of interesting observations one can additionaly make on 
the basis of the Leigh--Strassler constraint as this is given in equation (\ref{Fullcondition}).
Notice first that (\ref{Fullcondition}) reduces in the planar limit to:
\be
|h|^2 \left[\frac{1}{2} \left(|q|^2+\frac{1}{|q|^2}\right)
+\frac{1}{2}|\rho |^2 \right]
=|h|^2 \left[\cosh{(2\beta_{\mathbb{I}})}+\frac{1}{2}|\rho |^2\right] =g_{YM}^2
\ee   
This implies that when $\rho\neq 0$ the coupling constant h at the conformal fixed
point will be different from $g_{YM}$, in contrast to what happens for 
$\beta=\beta_{\mathbb{R}}\in\mathbb{R}$. In this sense, turning on $\rho$ is similar
to turning on the imaginary part of $\beta=\beta_{\mathbb{I}}$. Yet, there seems to
exist a particular point in the deformation space for which $h=g_{YM}$ continues
to hold in the large N limit. This occurs when:
\be\label{rhobeta}
\cosh{(2\beta_{\mathbb{I}})}+\frac{1}{2}|\rho|^2=1 \Rightarrow 
\beta_{\mathbb{I}}=\frac{1}{2}\arg\cosh{(1-\frac{|\rho|^2}{2})}
\ee
Closer inspection however of (\ref{rhobeta}) reveals that it has no possible
solutions, assuming $\beta_{\mathbb{I}}\in\mathbb{R}$ and $|\rho|>0$.  
This implies that despite appearances, there is no special point for which
$h=g_{YM}$ at two loops in the planar limit. Naturally, one expects that an analogous 
equation relating the two couplings, for which $h=g_{YM}$ at large N, may arise at 
any order in perturbation theory.  What is not clear of course, is whether
it will generically have any solutions or not. 

\section{Marginal deformations and Noncommutativity}

In \cite{Kulaxizi0610} we showed that for the $\beta$--deformed gauge theory 
it is possible to construct a noncommutativity matrix $\nc$ 
encoding in a precise manner information on the moduli space of the theory. 
This construction is very simple and is based on fundamental properties of the gauge 
theory and AdS/CFT. In what follows we will adopt the reasoning outlined in 
\cite{Kulaxizi0610} in order to determine a noncommutativity matrix for the 
$\rho$--deformation. We set $\beta=0$ for the time being and later on discuss
how to incorporate $\beta\neq 0$.

Our starting point is the F--term constraints:
\be\label{Fconstraints}
\begin{split}
\Phi_{1}\Phi_{2}&=\Phi_{2}\Phi_{1}+\rho \Phi_{3}^2,\quad
\Phi_{2}\Phi_{3}=\Phi_{3}\Phi_{2}+\rho \Phi_{1}^2,\quad
\Phi_{3}\Phi_{1}=\Phi_{1}\Phi_{3}+\rho \Phi_{2}^2\\
\ol{\Phi}_{1}\ol{\Phi}_{2}&=\ol{\Phi}_{2}\ol{\Phi}_{1}-\ol{\rho} \ol{\Phi}_{3}^2,\quad
\ol{\Phi}_{2}\ol{\Phi}_{3}=\ol{\Phi}_{3}\ol{\Phi}_{2}-\ol{\rho} \ol{\Phi}_{1}^2,\quad
\ol{\Phi}_{3}\ol{\Phi}_{1}=\ol{\Phi}_{1}\ol{\Phi}_{3}-\ol{\rho} \ol{\Phi}_{2}^2
\end{split}
\ee
from which we read the holomorphic and antiholomorphic parts of $\nc$
interpreting the eigenvalues of these matrices in the large N limit, as 
noncommuting coordinates parametrizing the space transverse to the worldvolume 
of the D3--brane. More precisely we have:
\be\label{holnc}
\begin{split}
[z_{1},z_{2}]&=\rho z_{3}^2,\quad [z_{2},z_{3}]=\rho z_{1}^2,\quad [z_{3},z_{1}]=\rho z_{2}^2\\
[\ol{z}_{1},\ol{z}_{2}]&=-\ol{\rho} \ol{z}_{3}^2,\quad [\ol{z}_{2},\ol{z}_{3}]=-\ol{\rho} \ol{z}_{1}^2,\quad [\ol{z}_{3},\ol{z}_{1}]=-\ol{\rho} \ol{z}_{2}^2\\
\end{split}
\ee
Following \cite{Kulaxizi0610} we would like to assume that there exists a star product
between some \emph{commuting} variables $z^{I},\ol{z}^{\bar{I}}$ which leads to 
commutation relations analogous to (\ref{holnc}), so that we can write for instance:
$i \nc^{12}=[z_{1},z_{2}]_{\ast}=z_{1}\ast z_{2}-z_{2}\ast z_{1}=\rho z_{3}^2$. This enables us to 
define a noncommutativity matrix which although position dependent, its entries are ordinary 
commuting objects. Then, under a change of coordinates $\nc$ will transform as a contravariant
antisymmetric tensor field.
We therefore write $\nc$ in matrix form as:
\be
\label{prencmatrix} 
\nc =
\begin{pmatrix} 
0&i\rho z_{3}^2&-i\rho z_{2}^2& \quad & \quad &\quad \\
-i\rho z_{3}^2&0&i\rho z_{1}^2 &\quad & \text{?} &\quad \\
i\rho z_{2}^2&-i\rho z_{1}^2&0 & \quad & \quad &\quad \\
\quad & \quad &\quad & 0&-i\ol{\rho}\bar{z}_{3}^2&i\ol{\rho}\bar{z}_{2}^2\\
\quad & \text{?} &\quad &i\ol{\rho}\bar{z}_{3}^2&0&-i\ol{\rho}\bar{z}_{1}^2\\
\quad & \quad &\quad & -i\ol{\rho}\bar{z}_{2}^2&i\ol{\rho}\bar{z}_{1}^2&0\\
\end{pmatrix}
\ee
It is clear that the F--term constraints determine the
(2,0) and (0,2) parts of $\nc$. D--terms will in principle specify the (1,1) piecies
of the noncommutativity matrix. However, as demonstrated in \cite{Kulaxizi0610}, 
there is an alternative indirect way of acquiring the information pertaining to D--terms. 
Recall that for the $\beta$--deformed gauge theory it was possible to fully 
determine $\nc$ by imposing certain simple conditions on its form --- namely
definite reality properties, symmetries and marginality. If there exists a choice for the $\nc^{I\bar{I}}$ components of the
noncommutativity matrix and the parameter $\rho$ which respects these 
requirements, we will be able to describe the deformation in noncommutative terms.
\footnote{Note for instance, that this description is not valid for the 
$\beta$--deformed theory when $\beta\in\mathbb{I}$.} 
We will see in the following that this is indeed the case here.

Let us first find out what are the possible (1,1) pieces of $\nc$
which respect the symmetries of the theory.
Consider for instance the commutator $[z^{1},\ol{z}^{\bar{2}}]=i \nc^{1\bar{2}}(z,\ol{z})$.
We easily see that: 
\begin{equation}\label{newa}
[z^{1},\ol{z}^{\bar{2}}]\xrightarrow{\mathbb{Z}_{3(2)}} \e^{-\frac{i 2\pi}{3}}[z^{1},\ol{z}^{\bar{2}}]\,.
\end{equation}
Eq. (\ref{newa}) constrains $\nc^{1\bar{2}}$ to either vanish or be a combination of any of the 
following: $\ol{z}^{\bar{1}}z^{3},\ol{z}^{\bar{3}}z^{2},z^{1}\ol{z}^{\bar{2}}$. All of the choices displayed are additionally invariant under the other discrete symmetry of the theory $\mathbb{Z}_{3(1)}$ as they should. 
%and that one of the choices coincides with the corresponding (1,1) piece of the noncommutativity 
%matrix for the $\beta$--deformed gauge theory. 
Several possibilities exist for the rest of the components of $\nc^{I\bar{J}}$ as well. In summary, the discrete global symmetries cannot completely fix the non-commutatvity matrix. To determine $\nc_\rho$ uniquely we transform $\nc$ to spherical coordinates 
\footnote{Refer to appendix \ref{nc} for the noncommutativity matrix in different coordinate systems.}. 
Then we require it to be real, transverse to and independent of the radial direction $r$. The last requirement
implements the exact maginality of the deformation in the dual description. 

Imposing these constraints we find that there are just \emph{two} distinct possibilities for
$\nc^{IJ}$. One of them is valid for $\rho\equiv -q_{1}\in \mathbb{R}$: 
\be\label{rhomatrixone} 
\nc_{1}=iq_{1}\begin{pmatrix} 0&
z_{3}^2&-z_{2}^2&0
&-z_{3}\bar{z}_{1}+z_{2} \bar{z}_{3}&z_{2}\bar{z}_{1}-z_{3}\bar{z}_{2}\\
-z_{3}^2&0& z_{1}^2&
\bar{z}_{2}z_{3}-z_{1}\bar{z}_{3}&0& -z_{1}\bar{z}_{2}+z_{3}\bar{z}_{1}\\
 z_{2}^2&-z_{1}^2&0&
-z_{2}\bar{z}_{3}+z_{1}\bar{z}_{2}&z_{1}\bar{z}_{3}-z_{2}\bar{z}_{1}&0\\
0&-z_{3}\bar{z}_{2}+z_{1}\bar{z}_{3}&z_{2}\bar{z}_{3}-z_{1}\bar{z}_{2}&0&
-\bar{z}_{3}^2&\bar{z}_{2}^2\\
z_{3}\bar{z}_{1}-z_{2}\bar{z}_{3}&0&-z_{1}\bar{z}_{3}+z_{2}\bar{z}_{1}&
\bar{z}_{3}^2&0&-\bar{z}_{1}^2\\
-z_{2}\bar{z}_{1}+z_{3}\bar{z}_{2}&z_{1}\bar{z}_{2}-z_{3}\bar{z}_{1}&0&
-\bar{z}_{2}^2&\bar{z}_{1}^2&0
\end{pmatrix}\ee
and the other one, for $\rho\equiv iq_{2}$ with $q_{2}\in \mathbb{R}$:
\be\label{rhomatrixtwo} 
\nc_{2}=q_{2}\begin{pmatrix} 0&
z_{3}^2&-z_{2}^2&0
&z_{3}\ol{z}_{1}+z_{2} \ol{z}_{3}&-z_{2}\ol{z}_{1}-z_{3}\ol{z}_{2}\\
-z_{3}^2&0& z_{1}^2&
-\ol{z}_{2}z_{3}-z_{1}\ol{z}_{3}&0& z_{1}\ol{z}_{2}+z_{3}\ol{z}_{1}\\
 z_{2}^2&-z_{1}^2&0&
z_{2}\ol{z}_{3}+z_{1}\ol{z}_{2}&-z_{1}\ol{z}_{3}-z_{2}\ol{z}_{1}&0\\
0&z_{3}\ol{z}_{2}+z_{1}\ol{z}_{3}&-z_{2}\ol{z}_{3}-z_{1}\ol{z}_{2}&0&
\ol{z}_{3}^2&-\ol{z}_{2}^2\\
-z_{3}\ol{z}_{1}-z_{2}\ol{z}_{3}&0&z_{1}\ol{z}_{3}+z_{2}\ol{z}_{1}&
-\ol{z}_{3}^2&0&\ol{z}_{1}^2\\
z_{2}\ol{z}_{1}+z_{3}\ol{z}_{2}&-z_{1}\ol{z}_{2}-z_{3}\ol{z}_{1}&0&
\ol{z}_{2}^2&-\ol{z}_{1}^2&0
\end{pmatrix}\ee
Combining the two into $\nc_{\rho}=\nc_{1}+\nc_{2}$ we define a unique noncommutativity 
matrix $\nc$ describing the $\rho$--deformation for general complex $\rho=(-q_{1}+iq_{2})\in\mathbb{C}$. 
This indicates that a noncommutative description of the transverse space is 
valid thoughout the whole of the $\rho$ parameter space, contrary to what happens for 
the $\beta$--deformed gauge theory.
 
Let us now discuss the properties of $\nc_\rho$.
Recall that the noncommutativity parameter for the $\beta$--deformed theory,
turned out to be position independent along isometry directions of the metric.
This was crucial for employing the Lunin--Maldacena generating technique.
We do not expect $\nc_\rho$ to be constant along isometry directions
since we know that the $\rho$--deformed theory does not respect any other global U(1)
symmetries except for the R--symmetry. Indeed, $\nc_\rho$ is of a highly nontrivial form even when 
written in spherical coordinates (see appendix \ref{nc}).
However, it would be nice to find a coordinate system for which $\nc_\rho$ is position 
independent, even if not along isometry directions \footnote{This is the case
for the nongeometric Q--space \cite{Sheltonetal,Ellwood:2006my,Loweetal}, for instance.}.

It is a curious fact that $\nc_\beta$ defined in (\ref{ncbetamatrix}) satisfies the following two conditions
\be\label{noncommutativityconditions}
\left.
\begin{aligned}
\text{Divergence Free Condition}\quad\quad\quad\quad\quad\quad\quad\quad\quad\quad\quad\quad\quad\quad\quad\p_{i}\nc^{ij}&=0\\
\text{Associativity Condition}\quad T^{[ijk]}\equiv \nc^{il}\p_{l}\nc^{jk}+\nc^{kl}\p_{l}\nc^{ij}+\nc^{jl}\p_{l}\nc^{ki}&=0 
\end{aligned}
\right\}
\quad \Rightarrow \quad T^{[ijk]}=\p_{l}(\nc^{l[i}\nc^{jk]})=0 
\ee 
%As discussed in \cite{Kulaxizi0610}, the possibility of finding a coordinate system in which a 
%non-commutativity matrix is constant may depend upon whether or not it satisfies eq. %(\ref{noncommutativityconditions}).
It is easy to see that $\nc_\rho$ satisfies the first condition but fails to preserve the associativity constraint.
This is disconcerting because it is not clear whether nonassociative deformations can be described through modified star products.
%an appropriate star product for the scalar fields a rather non--trivial one.
As a result it is far from obvious whether we can rewrite the Lagrangian of the $\rho$--deformed gauge theory as that of the $\Ncal=4$ Lagrangian with the usual product between the matter content of the theory replaced by some star product. Furthermore, a coordinate system in which $\nc_\rho$ is constant does not exist (contrary to what happens for the $\beta$--deformation)\footnote{This does not exclude the possibility of finding a reference frame for which $\nc_\rho$ is position independent. Integrability however will be lost.}. 

Finally, it is worth mentioning that the failure of associativity has its roots in the (1,1) parts of 
the noncommutativity matrix, thus challenging our method for determining them. 
There exists however a rather non--trivial check that we have constructed the correct $\nc$ describing the deformation.
We saw in the previous section, that for some special points in the space of couplings
of the marginally deformed theory, one can move from a theory where either $\beta$
or $\rho$ (but not both) is turned on, to a theory where both couplings are nonvanishing.
The whole analysis as well as the appropriate field redefinitions which took us from
one point to the other in the deformation space, relied on the holomorphicity of
the superpotential. It would thus appear quite improbable that we would be able to
see it happening in this context. In principle however, one would expect that if
the deformation is indeed described from an open string theory perspective as a 
noncommutative deformation of the transverse space, then at these special points 
$\nc$ should transform under a change of coordinates from $\nc_{\beta}$ or $\nc_{\rho}$ 
to $\nc=\nc_{\widetilde{\beta}}+\nc_{\widetilde{\rho}}$.
Moreover, one might hope that the coordinate transformation which would make this 
possible would be the precise analog of the field redefinition applied to
the gauge theory. 
Note however that in the case of the $\beta$--deformation, it is only for 
$\beta=\beta_{\mathbb{R}}\in\mathbb{R}$ that a noncommutative description --- with parameter 
$a=2\sin{\beta_{\mathbb{R}}}$ --- is valid.
This implies that we can apply the above consistency check if and only if
both the original and final points in the coupling constant space involve
a real parameter $\beta_{\mathbb{R}}$. A glance at the previous section will
convince us that this indeed occurs: starting with $\rho=q_{1}\in\mathbb{R}$
and $\beta=0$ one can reach a point with $\widetilde{\rho}=\frac{iq_{1}}{\sqrt{3}}\in\mathbb{I}$ 
and $\widetilde{a}=2\sin{\widetilde{\beta}}=\frac{2q_{1}}{\sqrt{3}}\in\mathbb{R}$ \footnote{We are here using
the result of eq. (18) approximated for $\beta\ll1$. The reason is that the non-commutativity matrix for the
$\beta$--deformation is valid only for small $\beta$ as shown in \cite{Kulaxizi0610} and at the end of section 2.1.}. 
In fact it is quite
straightforward to check that a coordinate transformation according to 
(\ref{Fieldredefinition2}) leads us from $\nc_{\rho}=-\nc_{1}$ to 
$\nc=\nc_{\widetilde{a}=\frac{2q_{1}}{\sqrt{3}}}+\nc_{\widetilde{\rho}=\frac{iq_{1}}{\sqrt{3}}}$. Furthermore, it 
appears that this case exhausts all possible coordinate changes that relate noncommutativity
matrices corresponding to different parameters of the Leigh--Strassler deformation.
We take this result as evidence that both our prescription
for determining the (1,1) parts of $\nc$ as well as the very interpretation of
the deformation in noncommutative terms are indeed justified.

%%%%%%%%%%%%%%%%%%%%%%%%%%%%%%%%%%%%%%%%%%%%%%%%%%%%%%%%%%%%%%%%%%%%%%%%%%%%%%%
\section{The Seiberg--Witten equations and the deformed flat space solution.}
%%%%%%%%%%%%%%%%%%%%%%%%%%%%%%%%%%%%%%%%%%%%%%%%%%%%%%%%%%%%%%%%%%%%%%%%%%%%%%%

In the previous section, we saw how the deformation of the superpotential affects the moduli space
of the gauge theory at large N. In particular, the six dimensional flat space with metric $\Gcal_{IJ}$
of the $\Ncal=4$ theory is promoted to a noncommutative space characterized now by the set 
$\Gcal_{IJ}$ and $\nc^{IJ}$.
Both metric and noncommutativity parameter are mainly determined from the Lagrangian of the theory; 
the former is read off from the kinetic term of the scalars while the latter from their potential.

Since an SU(N) gauge theory can be realized as the low energy limit of open strings attached on a 
stack of D3--branes, the set $(\Gcal_{flat},\nc)$ describes the geometry of the transverse space as 
seen by the open strings in the limit of large N and $\gra'\rar 0$. 
We will thus refer to $(\Gcal_{flat},\nc)$ as the \emph{open} string parameters .

On the other hand, any theory of open strings necessarily contains closed strings. 
Closed strings however perceive the geometry quite differently 
from open strings. In fact, it was shown in \cite{SeibergWitten9908, Schomerus} 
that target space noncommutativity from the point of view of open strings corresponds to turning 
on a B--field from the viewpoint of closed strings. The set $(g,B)$, with $g$ the closed string metric, 
are the \emph{closed} string parameters that describe the same geometry. In this context, $(g,B)$ 
represent the deformed flat space solution into which D3--branes are immersed 
\footnote{We are obviously interested here in the limit where open and closed strings are decoupled
from each other.}. 
Suppose now that we are given a set of equations relating the two groups of data. Then --- provided that
the open string parameters determined in the previous section exactly and fully describe the deformation --- 
we could specify the closed string fields $(g,B)$ of the deformed flat space geometry for free,
i.e. without having to solve the type IIB differential equations of motion \cite{Schwarz}.

Equations relating open and closed string parameters indeed exist in the literature \cite{Callanetal, Abouelsaoodetal, Schomerus, SeibergWitten9908}:
\be\label{SW}
\begin{split}
g&+B=\frac{1}{\Gcal^{-1}+\nc}\\ 
g_{s}&=G_{s} \sqrt{\frac{\det{\Gcal^{-1}}}{\det{(\Gcal^{-1}+\nc)}}}=G_{s} \sqrt{\frac{1}{\det{(1+\nc\Gcal)}}}\\
\end{split}
\ee
where $G_{s},g_{s}$ denote the corresponding open/closed string couplings
\footnote{Note that $G_{s}=1$ for the $\rho$--deformation.}. 
They were however considered in a situation somewhat different from the one discussed in this article, namely
for a flat D--brane embedded in flat background 
space with a constant B--field turned on along its worldvolume \cite{Schomerus, SeibergWitten9908}. 
It was under these circumstances that, the presence of the background B--field was shown to deform the algebra 
of functions on the worldvolume of the brane into that of a noncommutative Moyal type of algebra, where $\nc$
is a c--number.  
While it is natural to ask what happens in situations where the B--field is not constant, 
technical difficulties have hindered progress in this direction. In the order of increasing complexity, 
two cases can be considered: the case of a closed $\diff B=0$ though not necessarily constant two--form 
field B and the case of nonvanishing NS--NS three form flux $H=\diff B$ in a curved background. 
In \cite{CattaneoFelder99}
the former case was explored and the Moyal deformation of the algebra of functions on the
brane worldvolume was shown to naturally extended to the Kontsevich star product deformation 
\cite{Kontsevich97}. The authors of \cite{CornalbaSchiappa01} --- see also \cite{Herbstetal01,Herbstetal03,Ho} ---
undertook the study of the most general case where $H=\diff B\neq 0$. They considered a special
class of closed string backgrounds, called parallelizable, and expanded the background
fields in Taylor series. It was then possible to perturbatively analyze n--point
string amplitudes on the disk and obtain --- in a first order expansion ---
the appropriate generalization of (\ref{SW}). In fact, it turned out that equation (\ref{SW})
is still valid for a weakly varying nonclosed B--field even though the corresponding algebra
of functions is now both noncommutative and nonassociative.   

In this letter, we want to apply the above formulas in a situation where the B--field lies
in the \emph{transverse} space to the D3--brane. This case has
not been explicitly studied in the literature \footnote{Mainly because a constant B--field
in the transverse space can be gauged away leaving no trace on the geometry.} but one expects
by T--duality that equations (\ref{SW}) should continue to hold.
The most obvious concern here is that we do not have a set of conditions on the validity of (\ref{SW})
from the open string data. We have a non--commutative parameter which does not respect associativity and we have no way of knowing whether the corresponding B--field would be slowly varying or not.
Nevertheless, if the general reasoning is correct and (\ref{SW}) indeed provide the relation
between open and closed string parameters in this setup, the resulting closed string fields
$(g_{s},g,B)$ will constitute a new supergravity solution, i.e. the deformed flat space 
solution where D3--branes should be embedded. 

A natural place to test these thoughts first is the $\beta$--deformed theory for which
both the gravity dual and the corresponding deformed flat space solution
are known \cite{LuninMaldacena}. 
The open string data $(\Gcal_{flat},\nc_{\beta_{\mathbb{R}}})$ describing the $\beta$--deformation are
given in section 2. In this case the noncommutativity parameter $\nc_\beta$ turned out to be position independent although the associated NS--NS three form flux was non zero. 
%Can equations (\ref{SW}) reproduce the exact supergravity solution in this case? 
It is easy to show that applying (\ref{SW}) to the open string parameters 
$(\Gcal_{flat},\nc_{\beta_{\mathbb{R}}})$ one recovers the deformed flat space geometry found by 
Lunin and Maldacena in \cite{LuninMaldacena}.  This follows trivially from the fact that eq. (\ref{SW}) and the T--duality transformation rules of (\ref{Tduality}) are identical; yet the interpretation of the variables
involved is different. We will return to this point again in the following section. 
%(\ref{Tduality}) with which
%the solution was originally contstructed. We will return to this point in the 
%following section. 

To proceed, we check whether the closed string fields $(g_{s},g,B)$
determined from (\ref{SW}) for the $\rho$--deformation, satisfy the supergravity equations of motion.
It turns out that they do but only up to \emph{third} order in the deformation parameter $\rho$.
The discrepancy at higher orders is expected since there is no way to determine the validity of (\ref{SW}) for nonassociative deformations. At the same time, nonassociativity becomes manifest at second order in the deformation parameter. We postpone further discussion on this issue until section 6 and  
close this section with the explicit solution to this order.

\noindent The dilaton is given by
\be
\begin{split}
\mathrm{e}^{2 \Phi}&=G\\
G&=1+r_{1}^2\left[(q_{1}y-q_{2}x_{1})^2+(q_{1}y_{1}-q_{2}x)^2+(q_{1}x_{3}-q_{2}y_{2})^2+(q_{1}x_{2}-q_{2}y_{3})^2\right]+\\
&\quad +r_{2}^2\left[(q_{1}y-q_{2}x_{2})^2+(q_{1}y_{2}-q_{2}x)^2+(q_{1}x_{1}-q_{2}y_{3})^2+(q_{1}x_{3}-q_{2}y_{1})^2\right]+\\
&\quad +r_{3}^2\left[(q_{1}y-q_{2}x_{3})^2+(q_{1}x_{2}-q_{2}y_{1})^2+(q_{1}x_{1}-q_{2}y_{2})^2+(q_{1}y_{3}-q_{2}x)^2\right]
\end{split}
\ee
Here and in the following expressions, $\rho\equiv -2 q_1+i 2 q_2$ and $x,x_{i},y,y_{i}$ are defined as
\be\label{xidef}
\begin{split}
x_{1}&=-C_{1}r_{1}+C_{2}r_{2}+C_{3}r_{3}\quad 
x_{2}=C_{1}r_{1}-C_{2}r_{2}+C_{3}r_{3}\quad 
x_{3}=C_{1}r_{1}+C_{2}r_{2}-C_{3}r_{3}\\
y_{1}&=-S_{1}r_{1}+S_{2}r_{2}+S_{3}r_{3}\quad 
y_{2}=S_{1}r_{1}-S_{2}r_{2}+S_{3}r_{3}\quad 
y_{3}=S_{1}r_{1}+S_{2}r_{2}-S_{3}r_{3}\\
&\quad\quad x =C_{1}r_{1}+C_{2}r_{2}+C_{3}r_{3}\quad 
\quad\quad y =S_{1}r_{1}+S_{2}r_{2}+S_{3}r_{3}\\
\end{split}
\ee
where $S_{i},C_{i}$ represent the following triginometric functions:
\be\label{SCtrigdef}
\begin{split}
S_{1}&=\sin{(\vf_{2}+\vf_{3}-2\vf_{1})},\quad 
S_{2}=\sin{(\vf_{3}+\vf_{1}-2 \vf_{2})}, \quad
S_{3}=\sin{(\vf_{1}+\vf_{2}-2 \vf_{3})}\\
C_{1}&=\cos{(\vf_{2}+\vf_{3}-2 \vf_{1})},\quad
C_{2}=\cos{(\vf_{3}+\vf_{1}-2 \vf_{2})},\quad
C_{3}=\cos{(\vf_{1}+\vf_{2}-2 \vf_{3})}
\end{split} 
\ee
\noindent Using the definitions above, we can write the B--field as
\be
\begin{aligned}
B_{r_{1}r_{2}}&=r_{3}(q_{2}x_{3}-q_{1}y)\quad\quad\quad
B_{r_{2}r_{3}}=r_{1}(q_{2}x_{1}-q_{1}y)\quad\quad\quad
B_{r_{3}r_{1}}=r_{2}(q_{2}x_{2}-q_{1}y)\\
B_{r_{1}\vf_{2}}&=-r_{2}r_{3} (q_{1}x_{2}-q_{2}y_{1})\quad
B_{r_{1}\vf_{3}}=r_{2}r_{3}(q_{1}x_{3}-q_{2}y_{1})\quad
B_{r_{2}\vf_{1}}=r_{1}r_{3}(q_{1}x_{1}-q_{2}y_{2})\\
B_{r_{2}\vf_{3}}&=-r_{1}r_{3}(q_{1}x_{3}-q_{2}y_{2})\quad
B_{r_{3}\vf_{1}}=-r_{1}r_{2}(q_{1}x_{1}-q_{2}y_{3})\quad
B_{r_{3}\vf_{2}}=r_{1}r_{2}(q_{1}x_{2}-q_{2}y_{3})\\
B_{\vf_{1}\vf_{2}}&=r_{1}r_{2}r_{3}(q_{2}x-q_{1}y_{3})\quad
B_{\vf_{2}\vf_{3}}=r_{1}r_{2}r_{3}(q_{2}x-q_{1}y_{1})\quad
B_{\vf_{3}\vf_{1}}=r_{1}r_{2}r_{3}(q_{2}x-q_{1}y_{2})\,.\\
\end{aligned}
\ee
\noindent Finally, for the metric components we find the following complicated expressions 
\be
\begin{split}
g_{r_{1}r_{1}}&=1-
\left( r_{2}^{2} \left[ (q_{1}x_{3}-q_{2}y_{1})^{2}+(q_{1}y -q_{2}x_{2})^{2}\right]
+r_{3}^{2} \left[( q_{1}y-q_{2}x_{3})^{2}+(q_{1}x_{2}-q_{2}y_{1})^{2}   \right]  \right)\\
g_{r_{2}r_{2}}&=1-
\left(r_{1}^{2} \left[ (q_{1}x_{3}-q_{2}y_{2})^{2}+(q_{1}y -q_{2}x_{1})^{2}\right]
+r_{3}^{2} \left[( q_{1}y-q_{2}x_{3})^{2}+(q_{1}x_{1}-q_{2}y_{2})^{2}   \right]  \right)\\
g_{r_{3}r_{3}}&=1-
\left(r_{1}^{2} \left[ (q_{1}x_{2}-q_{2}y_{3})^{2}+(q_{1}y -q_{2}x_{1})^{2}\right]
+r_{2}^{2} \left[( q_{1}y-q_{2}x_{2})^{2}+(q_{1}x_{1}-q_{2}y_{3})^{2}   \right]  \right)\\
g_{\vf_{1}\vf_{1}}&=r_{1}^2 \left[1-
\left(r_{2}^{2} \left[ (q_{1}x_{1}-q_{2}y_{3})^{2}+(q_{1}y_{2} -q_{2}x)^{2}\right]
+r_{3}^{2} \left[( q_{1}y_{3}-q_{2}x)^{2}+(q_{1}x_{1}-q_{2}y_{2})^{2}   \right]  \right)\right]\\
g_{\vf_{2}\vf_{2}}&=r_{2}^2\left[1- 
\left(r_{1}^{2} \left[ (q_{1}x_{2}-q_{2}y_{3})^{2}+(q_{1}y_{1} -q_{2}x)^{2}\right]
+r_{3}^{2} \left[( q_{1}y_{3}-q_{2}x)^{2}+(q_{1}x_{2}-q_{2}y_{1})^{2}   \right]  \right)\right]\\
g_{\vf_{3}\vf_{3}}&=r_{3}^2 \left[1-
\left(r_{1}^{2} \left[ (q_{1}x_{3}-q_{2}y_{2})^{2}+(q_{1}y_{1} -q_{2}x)^{2}\right]
+r_{2}^{2} \left[( q_{1}y_{2}-q_{2}x)^{2}+(q_{1}x_{3}-q_{2}y_{1})^{2}   \right]  \right)\right]\\
g_{r_{1}r_{2}}&=r_{1}r_{2}\left[(q_{1}x_{3}-q_{2}y_{1})(q_{1}x_{3}-q_{2}y_{2})+(q_{1}y-q_{2}x_{1})(q_{1}y-q_{2}x_{2})\right]\\
g_{r_{1}r_{3}}&=r_{1}r_{3}\left[(q_{1}x_{2}-q_{2}y_{1})(q_{1}x_{2}-q_{2}y_{3})+(q_{1}y-q_{2}x_{3})(q_{1}y-q_{2}x_{1}) \right]\\
g_{r_{1}\vf_{1}}&=r_{1}\left(r_{2}^2\left[(q_{1}x_{3}-q_{2}y_{1})(q_{2}x-q_{1}y_{2})+(q_{2}x_{2}-q_{1}y)(q_{1}x_{1}-q_{2}y_{3})\right]+\right.\\
&\left. \quad +\quad r_{3}^2\left[(q_{2}x_{3}-q_{1}y)(q_{1}x_{1}-q_{2}y_{2})+(q_{1}x_{2}-q_{2}y_{1})(q_{2}x-q_{1}y_{3})\right]\right)\\
g_{r_{1}\vf_{2}}&=r_{1}r_{2}^2\left[(q_{2}x-q_{1}y_{1})(-q_{1}x_{3}+q_{2}y_{1})+(q_{2}x_{2}-q_{1}y)(-q_{1}x_{2}+q_{2}y_{3})\right]\\
g_{r_{1}\vf_{3}}&=r_{1}r_{3}^2\left[(q_{2}x-q_{1}y_{1})(-q_{1}x_{2}+q_{2}y_{1})+(q_{2}x_{3}-q_{1}y)(-q_{1}x_{3}+q_{2}y_{2})\right]\\
g_{r_{2}r_{3}}&=r_{2}r_{3}\left[(q_{1}x_{1}-q_{2}y_{3})(q_{1}x_{1}-q_{2}y_{2})+(q_{1}y-q_{2}x_{3})(q_{1}y-q_{2}x_{2})\right]\\
g_{r_{2}\vf_{1}}&=r_{2}r_{1}^2\left[(q_{2}x-q_{1}y_{2})(-q_{1}x_{3}+q_{2}y_{2})+(q_{2}x_{1}-q_{1}y)(-q_{1}x_{1}+q_{2}y_{3})\right]\\
g_{r_{2}\vf_{2}}&=r_{2}\left(r_{1}^2\left[(q_{1}x_{3}-q_{2}y_{2})(q_{2}x-q_{1}y_{1})+(q_{2}x_{1}-q_{1}y)(q_{1}x_{2}-q_{2}y_{3})\right]+\right.\\
&\left. \quad +\quad r_{3}^2\left[(q_{2}x_{3}-q_{1}y)(q_{1}x_{2}-q_{2}y_{1})+(q_{1}x_{1}-q_{2}y_{2})(q_{2}x-q_{1}y_{3})\right]\right)\\
g_{r_{2}\vf_{3}}&=r_{2}r_{3}^2\left[(q_{2}x_{3}-q_{1}y)(-q_{1}x_{3}+q_{2}y_{1})+(q_{2}x-q_{1}y_{2})(-q_{1}x_{1}+q_{2}y_{2})\right]\\
g_{r_{3}\vf_{1}}&=r_{3}r_{1}^2\left[(q_{2}x_{1}-q_{1}y)(-q_{1}x_{1}+q_{2}y_{2})+(q_{2}x-q_{1}y_{3})(-q_{1}x_{2}+q_{2}y_{3})\right]\\
g_{r_{3}\vf_{2}}&=r_{3}r_{2}^2 \left[(q_{2}x_{2}-q_{1}y)(-q_{1}x_{2}+q_{2}y_{1})+(q_{2}x-q_{1}y_{3})(-q_{1}x_{1}+q_{2}y_{3})\right]\\
g_{r_{3}\vf_{3}}&=r_{3}\left(r_{2}^2\left[(q_{1}x_{3}-q_{2}y_{1})(q_{2}x_{2}-q_{1}y)+(q_{2}x-q_{1}y_{2})(q_{1}x_{1}-q_{2}y_{3})\right]+\right.\\
&\left. \quad +\quad r_{1}^2\left[(q_{2}x_{1}-q_{1}y)(q_{1}x_{3}-q_{2}y_{2})+(q_{1}x_{2}-q_{2}y_{3})(q_{2}x-q_{1}y_{1})\right]\right)\\
g_{\vf_{1}\vf_{2}}&=r_{1}^2 r_{2}^2 \left[(q_{2}x-q_{1}y_{1})(-q_{1}x+q_{2}x)+(q_{2}y_{3}-q_{1}x_{2})(-q_{1}x_{1}+q_{2}y_{3})\right]\\
g_{\vf_{1}\vf_{3}}&=r_{1}^2 r_{3}^2 \left[(q_{2}x-q_{1}y_{1})(-q_{1}y_{3}+q_{2}x)+(q_{2}y_{2}-q_{1}x_{3})(-q_{1}x_{1}+q_{2}y_{2})\right]\\
g_{\vf_{2}\vf_{3}}&=r_{2}^2 r_{3}^2 \left[(q_{2}x-q_{1}y_{2})(-q_{1}y_{3}+q_{2}x)+(q_{2}y_{1}-q_{1}x_{3})(-q_{1}x_{2}+q_{2}y_{1})\right]\,.
\end{split}
\ee

%%%%%%%%%%%%%%%%%%%%%%%%%%%%%%%%%%%%%%%%%%%%%%%%%%%%%%%%%%%%%%%%%%
\section{D--branes in deformed \AdS and the near horizon geometry.}
%%%%%%%%%%%%%%%%%%%%%%%%%%%%%%%%%%%%%%%%%%%%%%%%%%%%%%%%%%%%%%%%%%%%

In this section we proceed to determine the gravity dual of 
the $\rho$--deformed gauge theory up to third order in the deformation parameter. 
Let us first discuss what happens in the case of the $\beta$--deformation where the dual geometry is known. In the previous section we observed that the T--duality transformation rules (\ref{Tduality}) with which the Lunin--Maldacena solution was constructed, are identical in form to (\ref{SW}).  To obtain the dual geometry for the $\beta$--deformation we saw in section 2 that we must use:
\be
E_{0}=g_{\mathrm{AdS}_{5}\times \Srm^5} \quad\text{and}\quad \Gamma=\nc_{\beta_{\mathbb{R}}}
\ee
Suppose now that we want to interpret these variables according to (\ref{SW}).
We would think of $g_{\mathrm{AdS}_{5}\times \Srm^5}$ as the open string
metric $\Gcal_{\mathrm{AdS}_{5}\times\Srm^{5}}$ whereas of $\Gamma$ as $\nc_{\beta_{\mathbb{R}}}$. In this sense, $(G_{s}=g_{YM}^2,\Gcal_{\mathrm{AdS}_{5}\times\Srm^{5}},\nc_{\beta_{\mathbb{R}}})$ would
encode the geometry as seen at large N by the open strings attached on a D3--brane embedded  
in the Lunin--Maldacena (\ref{LMAdS}) background. In the following we denote the NS--NS fields of the
solution as $(\widetilde{g}_{s},\widetilde{g},\widetilde{B})$.

In other words, consider a stack of N D3--branes in the deformed flat
space geometry of (\ref{LMflat}). The near horizon limit of this configuration is
the gravity dual of the Leigh--Strassler marginal deformation with 
$\beta=\beta_{\mathbb{R}}\in\mathbb{R}$ and $\rho=0$. A probe D3--brane propagating near the
stack will then be described by the DBI action written either in terms of the closed 
$(\widetilde{g}_{s},\widetilde{g},\widetilde{B})$ or of the open 
$(G_{s},\Gcal_{\mathrm{AdS}_{5}\times\Srm^{5}},\nc_{\beta})$ string fields. However, the action of a single D3--brane 
seperated from a collection of (N-1) other branes can also be obtained by integrating out the massive open
strings stretched between the probe and the source. 
Indeed, as expected according to \cite{Maldacena9705,Maldacena9709,Tseytlin:1999dj, MetsaevTseytlin,ChepelevTseytlin,Balasubramanianetal}, the DBI 
action describing the motion of a D3--brane in this background should
in the large N limit coincide with the leading IR part 
of the quantum effective action of the $\beta$--deformed theory 
obtained by keeping the U(1) external fields and integrating over the massive ones.

%It follows then that, in the large N limit, the DBI 
%action describing the motion of the D3--brane in this background would coincide with the leading IR part 
%of the quantum effective action of the $\beta$--deformed theory \cite{Maldacena9705,Maldacena9709,Tseytlin:1999dj, MetsaevTseytlin,ChepelevTseytlin,Balasubramanianetal}. This is the effective 
%action obtained by keeping the U(1) external fields and integrating over the massive ones 
%\footnote{When $\nc=0$ or for that matter, when $B=0$, we immediately recover the analogous $\Ncal=4$ SYM
%story, with the metric seen by both the open and the closed strings, being \AdS.}.

In this spirit, it does not seem surprising that the appropriate open string data 
%for the effective action of the $\beta$--deformed theory 
are the metric of \AdS and the noncommutativity parameter $\nc_{\beta_{\mathbb{R}}}$.
In fact, the action of the $\beta$--deformed gauge theory can be written as that of the parent
$\Ncal=4$ theory with the product of the matter fields replaced by a star product
associated to $\nc_{\beta_{\mathbb{R}}}$. Moreover, as conjectured in \cite{LuninMaldacena,Maurietal05,KuzenkoTseytlin} 
and later proven in \cite{Khoze}, all planar amplitudes are equal to their $\Ncal=4$ counterparts up to an 
overall phase factor. This suggests that the iterative structure of the large N $\beta$--deformed gauge 
theory amplitudes, when $\beta=\beta_{\mathbb{R}}\in\mathbb{R}$, is identical to that of the 
$\Ncal=4$ SYM theory. It is then not hard to imagine that the quantum effective action mentioned
above will be analogous to that of the undeformed theory with the only difference
being some phase factors coming from the noncommutative deformation of the product.
Subsequently, the open string fields appearing in the DBI form of the effective action of 
the $\Ncal=4$ theory $(G_{s},\Gcal,\nc=0)$ will be promoted to $(G_{s},\Gcal,\nc_{\beta_{\mathbb{R}}})$.

It is natural to wonder whether a similar situation could apply to the $\rho$--deformation as well. 
The results of section 3 suggest that this is likely not the case. Suppose we succeeded in writing the action of the $\rho$--deformed theory as the $\Ncal=4$ action
with a star product between the matter fields. It would still be difficult to understand how planar equivalence between the two theories would be achieved given that the deformation is both noncommutative and nonassociative.  
In fact, the proof given in \cite{Khoze} specifically relied on the associativity of the star product
for the $\beta$--deformation. However, nonassociativity shows up at second order in $\rho$ and in view of
the results of the previous section one may hope that a solution to this order could be obtained here too. 

To explicitly check if this is the case, we can use the second order expansion of (\ref{SW}):
\be\label{SW2}
\begin{split}
g&=\Gcal+\Gcal\nc\Gcal\nc\Gcal+\Ocal(\rho^4)\\
B&=-\Gcal\nc\Gcal+\Ocal(\rho^3)\\
G^{-1}&=1+\Tr \left[\Gcal\nc-\frac{1}{2}\Gcal\nc\Gcal\nc\right]+\Ocal(\rho^4)\,,\\
\end{split}
\ee
where $\Gcal$ is here the metric of $\mathrm{AdS}_{5}\times \Srm^5$ and $\Theta=\Theta_\rho$ defined in section 3.
Eqs. (\ref{SW2}) relate the open string parameters of the deformed theory
with the NS--NS string fields of the dual geometry. To find the RR--fluxes we resort to the type IIB equations of motion. We refer the reader to the appendix for the necessary definitions of the parameters involved as well as the type IIB field equations \cite{Schwarz} in five dimensions.
 %the background, 
%provide us only with the NS--NS background fields in a case where we expect that
%the RR-fields will also be non--trivial. 

We assume that there is no warp factor in front the metric to this order and make the standard ansatz for the five form field strength
\be\label{metricF5}
\begin{split}
\diff s_{10}^2&=\diff s_{AdS_{5}}^2+\diff s_{\widetilde{\Srm}^{5}}^2 \\
F_{5}&=f(\omega_{AdS_{5}}+\omega_{\widetilde{S}^{5}})  \,.  
\end{split}
\ee
Here $f$ is the normalization coefficient for the flux equal to $f=16\pi\mathrm{N}$, and $\omega_{AdS_{5}},\omega_{\widetilde{\Srm}^{5}}$
are the volume elements of the corresponding parts of the $\mathrm{AdS}_{5}\times\widetilde{\Srm}^{5}$
geometry.
Eq. (\ref{metricF5}) allows us to solve for the RR three form flux $F_{3}$:
\be\label{F3}
\begin{split} 
F_{3}&=-f^{-1} \diff\star_{5} e^{-2\Phi} H_{3}\\
H_{3}&=f^{-1}\diff\star_{5} F_{3}\Rightarrow \diff\left[B-f^{-1}\star_{5} F_{3}\right]=0\,.\\
\end{split} 
\ee
Note that to this order $F_{3}=f \star_{5} B$ which greatly simplifies calculations. 
This relation is mainly due to $\delta_{\Srm^{5}} \Omega_{\rho}=0$ where $\Omega_{\rho}=\Gcal\nc\Gcal$ denotes the form on $\Srm^{5}$ associated to the bivector $\nc_{\rho}$. 
It is clear from (\ref{SW2}) that $B=-\Omega$ to this order.
It is worth remarking that $F_{3}=f \star B$ is exact for the $\beta$--deformed theory where both 
$d\Omega_{\beta}=0$ and $\delta_{\Srm^{5}} \Omega_{\beta}=0$ hold. 
It is then possible to show that the type IIB equations are simultaneously satisfied up to third
order in the deformation parameter, for the following set of fields \footnote{We set $R=1$ where $R$ the radius of $\mathrm{A}\mathrm{d}\mathrm{S}_{5}$.}: 

\noindent The $F_{3}$ and $F_{5}$--form flux are simply given by
\be
\begin{split}
F_{3}&=\star_{S^{5}} \Omega  \quad\text{with}\quad \Omega\equiv \Gcal_{ik}\Gcal_{jl}\nc^{kl} dx^{i}\wedge dx^{j}\\
F_{5}&=f(\omega_{AdS_{5}}+G \omega_{\Srm^{5}})\,,\\
\end{split}
\ee
with $\Gcal$ the metric of $\mathrm{A}\mathrm{d}\mathrm{S}_{5}\times \mathrm{S}_{5}$.
\noindent The dilaton on the other hand can be expressed as follows
\be\label{DilatonDual}
\begin{split}
\mathrm{e}^{2\Phi}&=\mathrm{e}^{2\Phi_{0}} G\\
G^{-1}&=1+q_{1}^2 \left(v^2+\sa^2 u_{1}^2+(\ca^2+ \sa^2\ct^2 )u_{2}^2+(\ca^2+\sa^2\st^2 )u_{3}^2+\ca^2 v_{1}^2+\sa^2\st^2 v_{2}^2+\sa^2\ct^2 v_{3}^2\right)+\\
&+2 q_{1}q_{2}\left(u v+(\ca^2-\sa^2)u_{1}v_{1}-\left(\ct^2+\st^2 (\ca^2-\sa^2)\right)u_{2}v_{2}-\left(\st^2+\ct^2 (\ca^2-\sa^2)\right)\right)\\
&+q_{2}^2 \left(u^2+\ca^2 u_{1}^2+\sa^2\st^2 u_{2}^2+\sa^2\ct^2 u_{3}^2+\sa^2 v_{1}^2+\left(\ca^2+\sa^2\ct^2\right) v_{2}^2+\left(\ca^2+\sa^2\st^2\right) v_{3}^2\right)\,,\\
\end{split}
\ee
where to keep the expression compact we defined
\be\label{uvidef}
\begin{split}
u_{1}&=(-\ca C_{1}+\sa \st C_{2}+\sa\ct C_{3}),\,\,\,
u_{2}=(\ca C_{1}-\sa \st C_{2}+\sa\ct C_{3}),\,\,\,
u_{3}=(\ca C_{1}+\sa \st C_{2}-\sa\ct C_{3})\\
v_{1}&=(-\ca S_{1}+\sa \st S_{2}+\sa\ct S_{3}),\,\,\,
v_{2}=(\ca S_{1}-\sa \st S_{2}+\sa\ct S_{3}),\,\,\,
v_{3}=(\ca S_{1}+\sa \st S_{2}-\sa\ct S_{3})\\
&\quad \quad\quad\quad 
v=(\ca S_{1}+\sa \st S_{2}+\sa\ct S_{3})\quad \quad \quad \quad 
u=(\ca C_{1}+\sa \st C_{2}+\sa\ct C_{3})
\end{split}
\ee
with $(S_i,\,C_i)$ the trigonometric functions defined in (\ref{SCtrigdef}) and $(\ca,\sa,\ct,\st)\equiv (\cos{\alpha},\sin{\alpha},\cos{\theta},\sin{\theta})$ so that the parametrization of the deformed five-sphere is given in terms of the angular variables $(\alpha,\theta,\phi_1,\phi_2,\phi_3)$.
\noindent Using the same notations we write the components of the B--field as
\be
\begin{split}
&B_{\gra\theta}=\sa \left(q_{1} v+ q_{2} u\right)\quad
B_{\gra\vf_{1}}=0\quad
B_{\gra\vf_{2}}=\sa\st\ct \left(q_{1}u_{2}-q_{2}v_{2}\right)\quad
B_{\gra\vf_{3}}=\sa\st\ct \left(-q_{1}u_{3}+q_{2}v_{3}\right)\\
&B_{\theta\vf_{1}}=\ca\sa^2 \left(q_{1}u_{1}-q_{2}v_{1}\right)\quad
B_{\theta\vf_{2}}=-\ca\sa^2\st^2 \left(q_{1}u_{2}-q_{2}v_{2}\right)\quad
B_{\theta\vf_{3}}=-\ca\sa^2\ct^2 \left(q_{1}u_{3}-q_{2}v_{3}\right)\\
&B_{\vf_{1}\vf_{2}}=-\ca\sa^2\st\ct \left(q_{1}v_{3}+q_{2}u_{3}\right)\quad
B_{\vf_{2}\vf_{3}}=-\ca\sa^2\st\ct \left(q_{1}v_{1}+q_{2}u_{1}\right)\quad
B_{\vf_{3}\vf_{1}}=-\ca\sa^2\st\ct \left(q_{1}v_{2}+q_{2}u_{2}\right)\,.\\
\end{split}
\ee
\noindent Finally, we give the components of the metric $g$ on the deformed sphere:
\begin{equation}
\begin{split}
&g_{\gra\gra}=1-q_{1}^2 \left(\ct^2 u_{2}^2+\st^2 u_{3}^2+v^2\right)+
2 q_{1} q_{2}\left(-uv+\ct^2 u_{2}v_{2}+\st^2 u_{3}v_{3}\right)-
q_{2}^2\left(u^2+\ct^2 v_{2}^2+\st^2 v_{3}^2\right)\\
&g_{\theta\theta}=\sa^2\left[1-q_{1}^2\left(v^2+\sa^2 u_{1}^2+\ca^2 \left(\st^2 u_{2}^2+\ct^2 u_{3}^2\right)\right)-
2q_{1}q_{2}\left( uv-\sa^2 u_{1}v_{1}-\ca^2\left(\st^2 u_{2}v_{2}+\ct^2 u_{3}v_{3}\right)\right)-\right.\\
&\left. \quad\quad -q_{2}^2\left(u^2+\sa^2 v_{1}^2+\ca^2 \left(\st^2 v_{2}^2+\ct^2 v_{3}^2\right)\right)\right]\\
&g_{\vf_{1}\vf_{1}}=\ca^2\left[1-\sa^2\left(q_{1}^2\left(u_{1}^2+\st^2 v_{2}^2+\ct^2 v_{3}^2\right)+
2 q_{1}q_{2}\left(-u_{1}v_{1}+\st^2 u_{2}v_{2}+\ct^2 u_{3}v_{3}\right)+
q_{2}^2\left(v_{1}^2+\st^2 u_{2}^2+\ct^2 u_{3}^2\right)\right)\right]\\
&g_{\vf_{2}\vf_{2}}=\sa^2 \st^2 \left[1-q_{1}^2 \left( u_{2}^2 (\ca^2 +\sa^2\ct^2 )+\ca^2 v_{1}^2+ \sa^2\ct^2 v_{3}\right)+
2q_{1}q_{2}\left(-\ca^2 u_{1}v_{1}+(\ca^2 + \sa^2\ct^2 )u_{2}v_{2}-\sa^2 \ct^2 u_{3}^2 v_{3}^2\right)-\right.\\
&\left. \quad\quad -q_{2}^2\left(\ca^2 u_{1}^2+\sa^2\ct^2 u_{3}^2+(\ca^2+\sa^2\ct^2 )v_{2}^2\right)\right]\\
&g_{\vf_{3}\vf_{3}}=\sa^2\ct^2 \left[1-q_{1}^2\left(\ca^2 v_{1}^2+\sa^2\st^2 v_{2}^2+\left(\ca^2 +\sa^2\st^2\right) u_{2}^2\right)+
2q_{1}q_{2}\left(-\ca^2 u_{1}v_{1}-\sa^2\st^2 u_{2}v_{2}+\left(\ca^2+\sa^2\st^2\right) u_{3}v_{3}\right)-\right.\\
&\left. \quad\quad -q_{2}^2\left(\ca^2 u_{1}^2+\sa^2\st^2 u_{2}^2+\left(\ca^2 +\sa^2\st^2\right) v_{3}^2\right) \right]\\
&g_{\gra\theta}=\ca\sa\ct\st \left[q_{1}^2 \left( u_{2}^2-u_{3}^2\right)+ 
2 q_{1}q_{2}\left( -u_{2}v_{2}+u_{3}v_{3}\right)+
q_{2}^2\left( v_{2}^2-v_{3}^2\right)\right]\\
&g_{\gra\vf_{1}}=\ca\sa \left[q_{1}^2 \left( u_{1}v+ \st^2 u_{3}v_{2} \ct^2 u_{2}v_{3} \right)+
q_{1}q_{2} \left( u u_{1}-v v_{1}+u_{2}u_{3}-v_{2}v_{3})\right)-
q_{2}^2 \left( u v_{1}+\ct^2 u_{3}v_{2}+\st^2 u_{2}v_{3}\right) \right]\\
&g_{\gra\vf_{2}}=\ca\sa\st^2 \left[-q_{1}^2(u_{2}v+u_{3}v_{1})+
q_{1}q_{2}(-u u_{2}-u_{1}u_{3}+v v_{2}+v_{1}v_{3})+
q_{2}^2(u v_{2}+u_{1}v_{3}) \right]\\
&g_{\gra\vf_{3}}=\ca\sa\st^2 \left[-q_{1}^2(u_{2}v+u_{3}v_{1})+
q_{1}q_{2}(-u u_{2}-u_{1}u_{3}+v v_{2}+v_{1}v_{3})+
q_{2}^2(u v_{2}+u_{1}v_{3})\right]\\
&g_{\theta\vf_{1}}=\ca^2\sa^2\ct\st (q_{1}^2+q_{2}^2)(u_{3}v_{2}-u_{2}v_{3})\\ 
&g_{\theta\vf_{2}}=\sa^2\st\ct\left[q_{2}^2\left( u v_{2}+\sa^2 u_{3}v_{1}+\ca^2 u_{1}v_{3}\right)-
q_{1}q_{2}\left( u u_{2}-v v_{2}+u_{1}u_{3}-v_{1}v_{3}\right)
-q_{1}^2\left( u_{1}v+\ca^2 u_{3}v_{1}+\sa^2 u_{1}v_{3}\right)\right]\\
&g_{\theta\vf_{3}}=\sa^2\st\ct \left[q_{1}^2\left( u_{3}v+\ca^2 u_{2}v_{1}+\sa^2 u_{1}v_{2}\right)+
q_{1}q_{2}\left(u_{1}u_{2}-v_{1}v_{2}+uu_{3}-vv_{3}\right)-
q_{2}^2\left( u v_{3}+\sa^2 u_{2}v_{1}+\ca^2 u_{1}v_{2}\right)\right]\\
&g_{\vf_{1}\vf_{2}}=\ca^2\sa^2\st^2 (q_{1}^2+q_{2}^2) (u_{1}u_{2}+v_{1}v_{2})\\ 
&g_{\vf_{2}\vf_{3}}=\ca^2\sa^2\ct^2 (q_{1}^2+q_{2}^2) (u_{1}u_{3}+v_{1}v_{3})\\
&g_{\vf_{3}\vf_{1}}=\sa^4\ct^2\st^2 (q_{1}^2+q_{2}^2) (u_{2}u_{3}+v_{2}v_{3})\,.\\
\end{split}
\end{equation}

%Note that the definitions of these parameters is slightly different from the one used in the previous section.

%%%%%%%%%%%%%%%%%%%%%%%%%%%%%
\section{Discussion}
%%%%%%%%%%%%%%%%%%%%%%%%%%%%

In this article we studied the Leigh--Strassler marginal deformation of $\Ncal=4$ SYM
for $\rho\neq 0$. We made precise the relation of the deformation to noncommutativity by constructing
a noncommutativity matrix $\nc_\rho$ which describes it. We then considered $\nc_\rho$ as part of the open string data pertaining to the theory and used the Seiberg-Witten relations to obtain the corresponding closed string data. We were thus able to find supergravity solutions corresponding to the flat space deformation and the AdS/CFT dual of the deformed theory, up to third order in the deformation parameter.

\begin{figure}[hb]
  \centering
  \includegraphics[width=3in, height=2in]{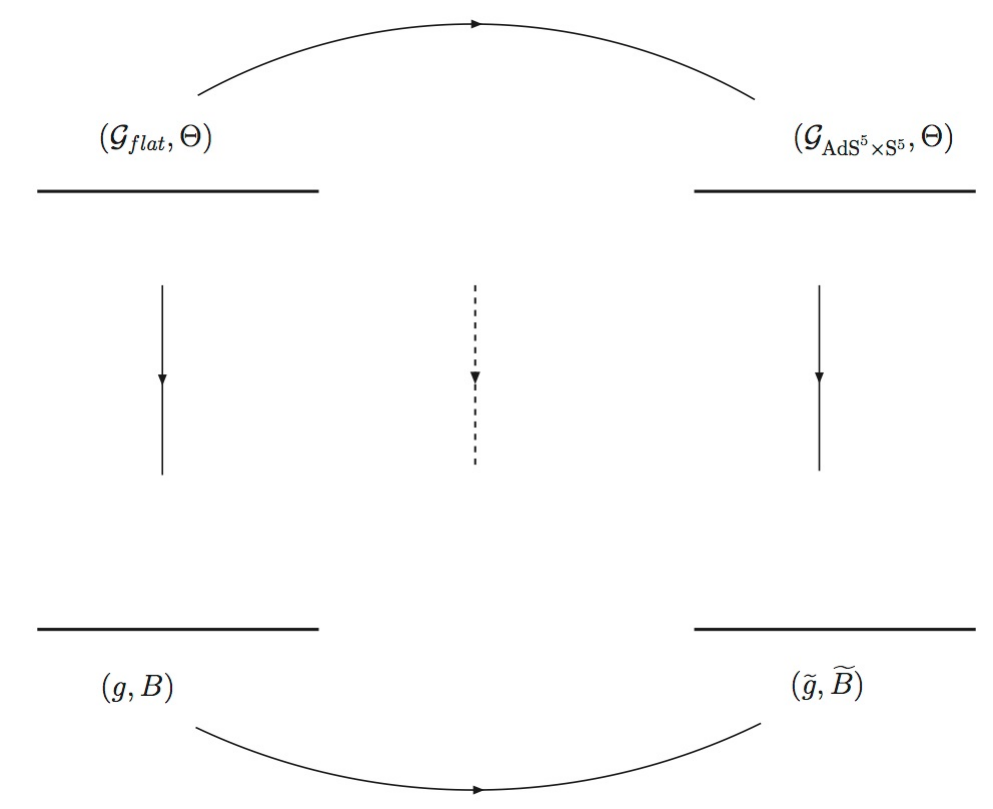}
  \caption
   {The mapping between open and closed string parameters.}
\end{figure}

\noindent The noncommutativity matrix $\nc_\rho$ is crucially different from $\nc_\beta$, the noncommutativity matrix describing the $\beta$--deformation, in its failure to preserve the property of associativity. The lack of associativity makes the possibility of defining a star product dubious. As a result, the Lagrangian of the $\rho$--deformed theory cannot be readily expressed in terms of the $\Ncal=4$ SYM Lagrangian with a modified product between the matter fields. 

%In fact, our efforts to construct appropriate star products produced extra terms in the action, to second and 
%third order in the deformation parameter $\rho$. Curiously enough, the
%additional terms in the Lagrangain were of the same form as the ones coming from 
%the $\beta$ and $\rho$ deformation themselves. We hope to explore this point further in the future. 

Similar issues arise when one considers the D--terms of the potential. It is possible to rewrite the D--terms of the $\Ncal=4$ theory as a sum of the F--terms plus an additional potential term involving the commutator between holomorphic and antiholomorphic matter fields:
\be\label{Dterms}
\Tr[\Phi_{I},\tilde{\Phi}^{I}][\Phi_{J},\tilde{\Phi}^{J}]=
\Tr[\Phi_{I},\Phi_{J}][\tilde{\Phi}^{I},\tilde{\Phi}^{J}]+
\Tr[\Phi_{I},\tilde{\Phi}^{J}][\Phi_{J},\tilde{\Phi}^{I}]
\ee
It is clear from (\ref{Dterms}) that should we wish to deform only the F--terms of the potential, 
we must appropriately alter the commutator: $[\Phi_{I},\tilde{\Phi}^{J}]$.
For the $\beta$--deformed gauge theory, the $(1,1)$ pieces of the 
noncommutativity matrix ensured that the D--terms remained unaffected 
by the deformation according to (\ref{Dterms}). The lack of a star product in the  
case of the $\rho$--deformation however, makes it impossible to perform this 
consistency check. 
  
Regarding the mapping between open and closed string fields; it is clear 
that eqs. (\ref{SW}) in sections 4 and 5 are not valid in this case,
especially due to the nonassociativity of $\nc_\rho$. 
It seems natural to expect that when $T^{ijk}$ of (\ref{noncommutativityconditions})
is nonvanishing, both $\nc$ and $T=\nc\p\nc$ are necessary for defining the deformation.
A natural generalization of (\ref{SW}) would then relate $(\Gcal,\nc,T=\nc\p\nc)$ to $(g,B,H=dB)$ and perhaps provide the deformed flat space solution to all orders in the deformation. 

Even if finding the appropriate mapping between open and closed string fields might help obtain
the deformed flat space geometry, it would not necessarily solve the problem of finding the dual gravity background as well. It is possible that nonassociativity spoils the planar equivalence between the $\Ncal=4$
theory and its deformation. This would obviously be reflected on the form
of the quantum effective action and therefore of the DBI, 
making it difficult to determine the relevant open string data.

%It is also worth remarking that (\ref{SW}) provides a relation only between the
%NS--NS fields of the open and closed string backgrounds. Information
%pertaining to the RR--fields is however essential, especially for determining the
%dual gravitational solution. In fact, in section 6 we had to rely on a particular
%ansatz for the metric and the five form flux in order to fully specify the background. 
%In this light, it may seem plausible that a different ansatz --- a warp factor in
%front of the AdS part of the metric, in particular --- could grant us the
%solution to all orders in the deformation parameter. The presence of a non--trivial
%warp factor may actually be related to the deviation of the coupling constant $h$
%in (\ref{Superpotential}) from its original $g_{YM}$ value. Indeed, in the case of
%the $\beta$--deformation, a warp factor is absent from the solution when $\beta\in\mathbb{R}$ 
%and $h=g_{YM}$, while it is not when $\beta_{\mathbb{I}}\neq 0$ and the Leigh--Strassler 
%constraint indicates that $h\neq g_{YM}$. This fact therefore represents another possible 
%explanation as to why our method fails to give an exact solution 
%\footnote{Note that we examined the case of a warp factor of the form $G^{n}$ with $n\in\mathbb{R}$
%and $G$ of (\ref{DilatonDual}) but found no value of $n$ for which the supergravity equations were satisfied
%to third order in the deformation parameter.}. 

So far we have considered the Leigh--Strassler marginal deformation at the point $\beta=0$. However, quantum corrections will probably generate a $\beta$--like term since a symmetry argument does not prohibit it. In this sense it is important to 
incorporate a nonvanishing $\beta$ in our discussion. This is easy to do, 
provided that $\beta=\beta_{\mathbb{R}}\in\mathbb{R}$. We can define
$\nc=\nc_{\beta_{\mathbb{R}}}+\nc_{\rho}$ and follow the same steps as in section 4 and 5.
The result is straightforward but does not cure the problems which appear 
at higher orders in $\rho$. 
The case of generic $\beta\in\mathbb{C}$ is more interesting but also more difficult to
study. A noncommutative description of the deformation is not valid in this case
and one relies on the $SL(2,\mathbb{R})_{s}$ symmetry of the supergravity equations
of motion in order to construct the dual solution \cite{LuninMaldacena}. Consequently,
there is no obvious way to incorporate a complex $\beta$ in our method. 

The reason that makes the case of complex $\beta$ worthwhile to explore further, is that 
according to the analysis of section 2, there exist some special points in the deformation
space which can take us from a theory of generic $\beta$ and $\rho=0$, to a marginal deformation
where both $\widetilde{\rho}$ and $\widetilde{\beta}$ are non vanishing. Since the gravity
dual in the former case is known, investigating the solution at these points may provide 
useful information on how to extend our results to all orders in the deformation parameters.

There are various possibilities for future work which range from addressing the questions raised
above, to establishing a precise connection with generalized complex geometry
\cite{Minasianetal, Ellwood:2006ya}, extending the relations between open and closed string parameters to include RR--fields and generalizing the results of \cite{deBoeretal,OoguriVafa0302, BerkovitsSeiberg} to incorporate supersymmetry. We hope to discuss some of these issues in the future.

%%%%%%%%%%%%%%%%%%%%
\vskip 2mm
\subsection*{Note added in proof} Several papers have investigated the subject since this article appeared on the Arxiv in Dec. 2006.
The dual gravity solution has not been constructed but the relation between exactly marginal deformations and noncommutativity was explored further in \cite{Mansson:2008xv}. The authors of \cite{Mansson:2008xv} discussed noncommutativity in the context of quantum groups. In this article, the same noncommutativity matrix was derived from a totally different perspective. Other interesting work on the $\rho$--deformation includes \cite{Halmagyi:2008dr,Halmagyi:2007ft,Halmagyi:2009te} in relation to generalized complex geometry and \cite{Bork:2007bj,Madhu:2007ew,Mansson:2007sh,Bork:2007bj,Minahan:2011dd,Minahan:2011bi}  regarding integrability and finitiness (see also \cite{Ardehali:2014zfa} for some recent developments).

%several \cite{Bundzik:2005zg}

%%%%%%%%%%%%%%%%%%%%%%%%%%

\vskip 20mm
\subsubsection*{Acknowledgments}
I am indebted to A. Murugan and K. Zoubos for numerous discussions that greatly contributed to my understanding of several aspects of this work. I am also grateful to J. Maldacena and L. Rastelli for helpful conversations and encourangement as well as to J.P. Gauntlett, A. Naqvi, A. Parnachev, R. Ricci and M. Ro\v cek.   

%%%%%%%%%%%%%%%%%%%%%%%

%%%%%%%%%
\appendix
%%%%%%%%%%

\section{The noncommutativity matrix}\label{nc}

Here we present the noncommutativity matrix in polar coordinates $(r_{i},\vf_{i})$ with $i=1,2,3$ on 
$\mathbb{R}^6$. We assume that $\nc_{\rho}$ is given in terms of commuting variables $(z,\ol{z})$ and that 
we can follow the transformation rules of contravariant tensors when changing coordinate systems, namely:
\be\label{transformationrule}
\nc^{i'j'}=\frac{\p x'^{i'}}{\p x^{i}}\frac{\p x'^{j'}}{\p x^{j}}\nc^{ij}
\ee 
Rescalling $q_{i}$ of (\ref{rhomatrixone}) and (\ref{rhomatrixtwo}) as $q_{i} \rar 2 q_{i}$ then yields: 
\be\label{rhopolar}
\nc_{\rho}=
\left(\begin{smallmatrix}
0& -(q_{2}x_{3}-q_{1}y)r_{3}& (q_{2}x_{2}-q_{1}y) r_{2}& 0& \frac{(q_{1}x_{2}-q_{2}y_{1})r_{3}}{r_{2}} & \frac{(q_{1}x_{3}-q_{2}y_{1})r_{2}}{r_{3}}  \\
 (q_{2}x_{3}-q_{1}y) r_{3}& 0 & - (q_{2}x_{1}-q_{1}y) r_{1}&-\frac{(q_{1}x_{1}-q_{2}y_{2})r_{3}}{r_{1}} &0 &\frac{(q_{1}x_{3}-q_{2}y_{2})r_{1}}{r_{3}} \\
-(q_{2}x_{2}-q_{1}y)r_{2} &(q_{2}x_{1}-q_{1}y)r_{2} & 0& \frac{(q_{1}x_{1}-q_{2}y_{3})r_{2}}{r_{1}}&-\frac{(q_{1}x_{2}-q_{2}y_{3})r_{1}}{r_{2}} & 0 \\
 0&\frac{(q_{1}x_{1}-q_{2}y_{2})r_{3}}{r_{1}} &- \frac{(q_{1}x_{1}-q_{2}y_{3})r_{2}}{r_{1}}& 0&-\frac{(q2 x - q1 y3)r_{3}}{r_{1}r_{2}} & \frac{(q_{2}x-q_{1}y_{2})r_{2}}{r_{1}r_{3}}\\
 -\frac{(q_{1}x_{2}-q_{2}y_{1})r_{3}}{r_{2}}&0 &\frac{(q_{1}x_{2}-q_{2}y_{3})r_{1}}{r_{2}} & \frac{(q_{2}x-q_{1}y_{3})r_{3}}{r_{1}r_{2}}&0 &- \frac{(q_{2}x-q_{1}y_{1})r_{1}}{r_{2}r_{3}} \\
 \frac{(q_{1}x_{3}-q_{2}y_{1})r_{2}}{r_{3}}& -\frac{(q_{1}x_{3}-q_{2}y_{2})r_{1}}{r_{3}}& 0& -\frac{(q_{2}x-q_{1}y_{2})r_{2}}{r_{1}r_{3}}&\frac{(q_{2}x-q_{1}y_{1})r_{1}}{r_{2}r_{3}} & 0\\
\end{smallmatrix}\right)
\ee
where to keep the expressions compact, we defined variables $x,x_{i}$ and $y,y_{i}$ according to:
\be\label{nnA}
\begin{split}
x_{1}&=-C_{1}r_{1}+C_{2}r_{2}+C_{3}r_{3}\quad 
x_{2}=C_{1}r_{1}-C_{2}r_{2}+C_{3}r_{3}\quad 
x_{3}=C_{1}r_{1}+C_{2}r_{2}-C_{3}r_{3}\\
y_{1}&=-S_{1}r_{1}+S_{2}r_{2}+S_{3}r_{3}\quad 
y_{2}=S_{1}r_{1}-S_{2}r_{2}+S_{3}r_{3}\quad 
y_{3}=S_{1}r_{1}+S_{2}r_{2}-S_{3}r_{3}\\
&\quad\quad x =C_{1}r_{1}+C_{2}r_{2}+C_{3}r_{3}\quad 
\quad\quad y =S_{1}r_{1}+S_{2}r_{2}+S_{3}r_{3}\\
\end{split}
\ee
whereas $S_{i},C_{i}$ represent the following triginometric functions:
\be\label{mmA}
\begin{split}
S_{1}&=\sin{(\vf_{2}+\vf_{3}-2\vf_{1})},\quad 
S_{2}=\sin{(\vf_{3}+\vf_{1}-2 \vf_{2})}, \quad
S_{3}=\sin{(\vf_{1}+\vf_{2}-2 \vf_{3})}\\
C_{1}&=\cos{(\vf_{2}+\vf_{3}-2 \vf_{1})},\quad
C_{2}=\cos{(\vf_{3}+\vf_{1}-2 \vf_{2})},\quad
C_{3}=\cos{(\vf_{1}+\vf_{2}-2 \vf_{3})}
\end{split} 
\ee

The discrete symmetry $\mathbb{Z}_{3(1)}\times\mathbb{Z}_{3(2)}$ along with the $U(1)_{R}$ are particularly 
transparent in this form. \newline Observe first that under $\mathbb{Z}_{3(1)}$:
\be 
\begin{split}
\mathbb{Z}_{3(1)}: \quad 
\left(x_{1},x_{2},x_{3},y_{1},y_{2},y_{3}\right)& \rar\left(x_{3},x_{1},x_{2},y_{3},y_{1},y_{2}\right)\\
\text{while}\quad (x,y)& \rar (x,y)\\
\end{split}
\ee 
Then it is easy to see for example, that $\nc_{\rho}^{r_{1}\vf_{2}}=\frac{(q_{1}x_{2}-q_{2}y_{1})r_{3}}{r_{2}}\rar \nc_{\rho}^{r_{3}\vf_{1}}=\frac{(q_{1}x_{1}-q_{2}y_{3})r_{2}}{r_{1}}$. \newline
The action of $\mathbb{Z}_{3(2)}$ is equally simple transforming the polar angles $\vf_{i}$ 
as:
\be 
\mathbb{Z}_{3(2)}:\quad(\vf_{1},\vf_{2},\vf_{3})\rar (\vf_{1},\vf_{2}+\frac{2\pi}{3},\vf_{3}-\frac{2\pi}{3})
\ee 
thus leaving invariant the trigonometric functions $S_{i},C_{i}$ which depend on the following combinations: \newline
$\sigma_{i}\equiv\frac{1}{3}(\vf_{i+1}+\vf_{i+2}-2\vf_{1})$. 
Morever, note that $\nc_{\rho}$ is independent of $\psi=\frac{1}{3}(\vf_{1}+\vf_{2}+\vf_{3})$ therefore
respects the $U(1)_{R}$ R--symmetry of the theory.

In a similar manner, one obtains the noncommutativity matrix $\nc_{\rho}$ in spherical coordinates denoted as $(r,\gra,\theta,\vf_{1},\vf_{2},\vf_{3})$.
\be
\begin{split}
z_{1}&=r \cos{\alpha} \mathrm{e}^{i \phi_{1}}, \quad z_{2}=r \sin{\alpha}\sin{\theta} \mathrm{e}^{i \phi_{2}}, \quad
z_{3}=r \sin{\alpha}\cos{\theta} \mathrm{e}^{i \phi_{3}}\\
\ol{z}_{1}&=r \cos{\alpha} \mathrm{e}^{-i \phi_{1}}, \quad \ol{z}_{2}=r \sin{\alpha}\sin{\theta} \mathrm{e}^{-i \phi_{2}}, \quad \ol{z}_{3}=r \sin{\alpha}\cos{\theta} \mathrm{e}^{-i \phi_{3}}
\end{split}
\ee
where it reads\footnote{We use here the following abbreviations: $\sa=\sin{\gra}, \ca=\cos{\gra}, \st=\sin{\theta},\ct=\cos{\theta}$.}: 
\be\label{rhospherical}
\nc_{\rho}=
\begin{pmatrix}
0&-\frac{q_{2}u+q_{1}v}{\sa}&0&\frac{\ct(-q_{1}u_{2}+q_{2}v_{2})}{\sa\st}&\frac{\st(q_{1}u_{3}-q_{2}v_{3})}{\sa\ct}\\
\frac{q_{2}u+q_{1}v}{\sa}&0 &\frac{-q_{1}u_{1}+q_{2}v_{1}}{\ca} &\frac{\ca(q_{1}u_{2}-q_{2}v_{2})}{\sa^2} &\frac{\ca(q_{1}u_{3}-q_{2}u_{3})}{\sa^2} \\
0&\frac{q_{1}u_{1}-q_{2}v_{1}}{\ca} &0 &\frac{\ct(q_{2}u_{3}+q_{1}v_{3})}{\ca\ct} &-\frac{\st(q_{2}u_{2}+q_{1}v_{2})}{\ca\ct} \\
\frac{\ct(q_{1}u_{2}-q_{2}v_{2})}{\sa\st}&\frac{\ca(-q_{1}u_{2}+q_{2}v_{2})}{\sa^2} &-\frac{\ct(q_{2}u_{3}+q_{1}v_{3})}{\ca\ct} &0 &\frac{\ca(q_{2}u_{1}+q_{1}v_{1})}{\sa^2\st\ct} \\
-\frac{\st(q_{1}u_{3}-q_{2}v_{3})}{\sa\ct}& \frac{\ca(-q_{1}u_{3}+q_{2}v_{3})}{\sa^2}& \frac{\st(q_{2}u_{2}+q_{1}v_{2})}{\ca\ct}&-\frac{\ca(q_{2}u_{1}+q_{1}v_{1})}{\sa^2\st\ct} &0 \\
\end{pmatrix}
\ee
Note that $\nc_{\rho}$ is now a five--dimensional matrix along the $S^{5}$ and that variables $u,u_{i},v,v_{i}$ 
appearing in (\ref{rhospherical}) are defined as:
\be\label{nnB}
\begin{split}
u_{1}&=(-\ca C_{1}+\sa \st C_{2}+\sa\ct C_{3}),\quad
u_{2}=(\ca C_{1}-\sa \st C_{2}+\sa\ct C_{3}),\quad
u_{3}=(\ca C_{1}+\sa \st C_{2}-\sa\ct C_{3})\\
v_{1}&=(-\ca S_{1}+\sa \st S_{2}+\sa\ct S_{3}),\quad
v_{2}=(\ca S_{1}-\sa \st S_{2}+\sa\ct S_{3}),\quad
v_{3}=(\ca S_{1}+\sa \st S_{2}-\sa\ct S_{3})\\
&\quad \quad\quad\quad 
v=(\ca S_{1}+\sa \st S_{2}+\sa\ct S_{3})\quad \quad \quad \quad 
u=(\ca C_{1}+\sa \st C_{2}+\sa\ct C_{3})
\end{split}
\ee
It is then clear that $\nc_{\rho}$ is independent of the radial direction r.

\section{RR--fields and supergravity equations of motion}\label{SUGRA}

As mentioned previously, although the procedure proposed in this article
gives us the solution for the NS--NS fields of the geometry for free,
it does not produce any information on the RR--ones. We thus have to
compute them using the supergravity equations of motions \cite{Schwarz}.
We employ the following ansatz \footnote{Note that the vanishing axion condition can be deduced from the other 
two in equation (\ref{ansatz}).}:
\be\label{ansatz}
\begin{split}
\diff s_{10}^2&=\diff s_{AdS_{5}}^2+\diff s_{5}^2\\
C&=0 \quad F_{5}=f (\omega_{AdS_{5}}+\omega_{\widetilde{\Srm}^{5}})\\
\end{split}
\ee
where $f$ is the appropriate normalization coefficient for the flux
which in this case reduces to $f=16\pi\mathrm{N}$ and $\omega_{AdS_{5}},\omega_{\widetilde{\Srm}^{5}}$
are the volume elements of the corresponding parts of the $\mathrm{AdS}_{5}\times\widetilde{\Srm}^{5}$
geometry.
Then the supergravity field equations reduce to:
\be
\begin{split} 
D^2 e^{-2 \Phi}&=-\frac{1}{6} \left(F_{3}^2-e^{-2\Phi} H_{3}^2 \right)\\ 
F_{3}&=-f^{-1} \diff\star_{5} e^{-2\Phi} H_{3}\\
H_{3}&=f^{-1}\diff\star_{5} F_{3}\\
R_{MN}&=-2 D_{M}D_{N}\Phi-\frac{1}{4}g_{MN} D^2 \Phi+\frac{1}{2} g_{MN} \p_{R}\Phi\p^{R}\Phi+\\
&+\frac{1}{96} e^{2\Phi}F_{MPQR}F_{N}^{PQR}+\frac{1}{4}
(H_{MPQ}H_{N}^{PQ}+e^{2\Phi} F_{MPQ}F_{N}^{PQ})-\frac{1}{48}
g_{MN}(H_{3}^2+e^{2\Phi}F_{3}^2)
\end{split} 
\ee
where $M,N$ represent five dimensional indices on the compact piece of the geometry
whereas $\star_{5}$ denotes the Hodge star on the same manifold.

%%%%%%%%%%%%%%%%%%%%%%%%%%%%%%%%%%%%%%%%%%%%%%%%%%%%%%%%

\bibliography{md_refs}
\bibliographystyle{JHEP}

\end{document}